\documentclass[twocolumn]{aastex63}
\usepackage{bm}
\usepackage{threeparttable}
\usepackage{morefloats}
\usepackage{CJK}

\def\msun{M_\odot}

\def\<{\,\langle\langle}
\def\>{\,\rangle\rangle}

\newcommand\rg{r_{\mathrm{g}}}

\def\sixteen{\text{\rm AGNUV4}}
\def\twelve{\text{\rm AGNUV0.03}}
\def\sixteenB{\text{\rm AGNUVB0.6}}
\def\eighteenB{\text{\rm AGNUVB3}}

\shorttitle{UV/Optical Regions of AGN Disks}
\shortauthors{Jiang et al.}

\begin{document}
\begin{CJK*}{UTF8}{gbsn}

\title{Radiation and Magnetic Pressure Support in Accretion Disks around Supermassive Black Holes and The Physical Origin of the Extreme Ultraviolet to Soft X-ray Spectrum}

\correspondingauthor{Yan-Fei Jiang}
\email{yjiang@flatironinstitute.org}

\author[0000-0002-2624-3399]{Yan-Fei Jiang (姜燕飞)}
\affiliation{Center for Computational Astrophysics, Flatiron Institute, New York, NY 10010, USA}

\author{Omer Blaes}
\affiliation{Department of Physics, University of California, Santa Barbara, CA 93106, USA}

\author[0009-0001-1399-2622]{Ish Kaul}
\affiliation{Department of Physics, University of California, Santa Barbara, CA 93106, USA}

\author[0000-0003-0232-0879]{Lizhong Zhang (张力中)}
\affiliation{Center for Computational Astrophysics, Flatiron Institute, New York, NY 10010, USA}

\begin{abstract}
We present the results of four three-dimensional radiation magnetohydrodynamic simulations of accretion disks around a $10^8$ solar mass  black hole, which produce the far ultraviolet spectrum peak and demonstrate a robust physical mechanism to produce the extreme ultraviolet to soft X-ray power-law continuum component.
The disks are fed from rotating tori and reach accretion rates ranging from $0.03$ to $4$ times the Eddington value. The disks become radiation pressure or magnetic pressure dominated depending on the relative timescales of radiative cooling and gas inflow. Magnetic pressure supported disks can form 
with or without net poloidal magnetic fields as long as the inflowing gas can cool quickly enough, which can typically happen when the accretion rate is low.  We calculate the emerging spectra from these disks using multi-group radiation transport with realistic opacities and find that they typically peak around $10$ eV.  
At accretion rates close to or above the Eddington limit, a power-law component can appear for photon energies between $10$ eV and 1 keV with a spectral slope varying between $L_\nu\propto\nu^{-1}$ and $\nu^{-2}$, comparable to what is observed in radio quiet quasars.  The disk with $3\%$ Eddington accretion rate does not exhibit this component.  These high energy photons are produced in an optically thick region $\approx 30^{\circ}-45^{\circ}$ from the disk midplane 
by compressible bulk Comptonization within the converging accretion flow. Strongly magnetized disks that have a very small surface density will produce a spectrum that is very different from what is observed. 
\end{abstract}


\section{Introduction}

Active Galactic Nuclei (AGN) have long been suspected to be powered by the accretion of matter onto supermassive black holes \citep{LYN69,REES1984}, but the physical nature of the accretion flow continues to be very uncertain. 
Optical reverberation campaigns \citep{cackett2021} and microlensing observations \citep{Morganetal2010}
have revealed that the size scale of emission increases with rest wavelength in a manner that is consistent with accretion power being thermalized into radiation in an optically thick medium.  That medium is widely suspected to be a rotationally supported accretion flow, resembling an accretion disk, though the observed size scales are several times larger than predicted by a standard geometrically thin disk \citep{SS73}.  Optical-near infrared spectra \citep{kishimoto2008} are also consistent with a superposition of thermalized emission at different radii.  At shorter wavelengths, the observed discrepancies with standard accretion disk model spectra are more severe.  For example, ultraviolet spectra have a quasi-universal shape with a break to a power-law beyond 1000~\AA\ (12 eV) \citep{ZHE97}.  The universality of this shape is in strong disagreement with standard accretion disk theory (e.g. \citealt{LAO14,Mitchell+2023,Kynoch+2023,Jin+2024}).  In addition to spectra, observed variability time scales are far less than the inflow time scale of standard geometrically thin accretion disks (e.g. \citealt{DexterBegelman2019}).

One unique feature that shows up in many AGN spectra is the soft X-ray excess. Observationally, for type I AGNs, the luminosity in the energy range $\approx 0.2-2$ keV is typically higher than the value extrapolated downward from 
the power-law spectral shape of hard ($>2$ keV) X-rays \citep{Arnaud+1985,TurnerPounds1988,LAO97,GierlinskiDone2004,Done+2012}. 
The soft X-ray spectrum can be typically fitted by a blackbody component with temperature $\approx 0.1-0.2$ keV, which is almost independent of black hole mass and mass accretion rate \citep{GierlinskiDone2004,Bianchi+2009}. This temperature is much higher than the typical gas photospheric temperature in standard thin disks around supermassive black holes, which should also vary for different black hole masses and accretion rates. 
If this component is produced by thermal Comptonization \citep{CzernyElvis1987,Czernyetal2003,Jin+2012,Rozanska+2015,Petrucci+2018,Petrucci+2020}, it would
also require an electron temperature $0.1-0.2$ keV and relatively high optical depth $\sim 45$ \citep{GierlinskiDone2004,Palit+2024,Ballantyne+2024}. This would imply that the disk has two distinct thermal Comptonization 
components to produce the hard X-rays and the extreme ultraviolet to soft X-rays, 
with a significant fraction of energy dissipated in the latter.

One aspect that is in fact suggestive of some sort of Comptonization is that the soft X-ray excess may, in many systems, be a high energy extrapolation of the far ultraviolet spectrum beyond 1000~\AA\ \citep{LAO97,Czernyetal2003}.  Indeed, the continuum far UV and soft X-ray spectra of radio quiet quasars are
consistent in both slope and amplitude, suggesting that one can simply join them across the generally unobservable extreme ultraviolet region.  In particular, the slopes are $L_\nu\propto\nu^{-1.77}$ in the far UV \citep{ZHE97} and $L_\nu\propto\nu^{-1.72}$ in the soft X-rays \citep{LAO97}.  There is no analogous power-law component in X-ray binaries, suggesting that a 
fundamental difference exists between accretion disks in X-ray binaries and AGNs. 

Several models have been proposed to explain the origin of the soft X-ray excess. The almost constant fitted temperature has motivated ideas based on
reflection \citep{Crummyetal2006} or absorption \citep{GierlinskiDone2004,SobolewskaDone2007,BallantyneXiang2020,Xiang+2022} of hard X-rays combined with an abrupt increase in opacity between $0.7-3$ keV due to O VII, O VIII and Fe L transitions. However, these models typically require a very high smearing velocity ($\sim 30\% - 80\%$ of the speed of light) to hide the sharp atomic features, as the observed soft X-ray spectra are typically very smooth and broad. It is also typically very hard for reflection models to reproduce the observed high ratio of soft to hard X-ray luminosity 
\citep{SobolewskaDone2007,Middleton+2009}. Other models suggest that AGNs showing soft X-ray excesses typically have accretion rates close to or above the Eddington limit (particularly Narrow Line Seyfert 1s) and the gas temperature in these disks should be high enough to produce soft X-rays via thermal emission \citep{Mineshige+2000,WangNetzer2003}. However, it is hard to understand the almost constant temperature of the soft X-ray excess in this model, and the assumed model for super-Eddington accretion disks is quite different from what is found in recent numerical simulations \citep{JiangDai2024}. 


One important limitation of all these phenomenological models is the uncertainty of accretion disk structures around supermassive black holes. 
Recent advances in radiation magneto-hydrodynamic simulations show many interesting properties of the disks that cannot be captured by the classical accretion disk models \citep{Davis+2020,JiangDai2024}. Magnetic fields are known to play an important role in accretion disks since magneto-rotational instability (MRI) is believed to be the physical mechanism for angular momentum transport in the disk \citep{BAL91,BAL98}.  
However, local shearing box simulations also find that if the disk is supplied with enough vertical magnetic field, the disk can end up in a state where magnetic pressure mainly due to toroidal magnetic fields is much larger than the thermal pressure and vertical gravity is balanced by magnetic fields instead of thermal pressure gradients \citep{BAI13,Salvesen+2016a,Salvesen+2016b}. Global simulations with different levels of complexity also find similar results \citep{JohansenLevin2008,Gaburov+2012,Fragile2017,Mishra+2020}.  Without net vertical magnetic fields, initial super-thermal toroidal magnetic fields are needed to achieve magnetic pressure supported disks in order that the dynamo in the disk can generate enough magnetic field to balance buoyant field loss \citep{Squire+2024}.  
These simulation results partially confirm some analytical models of magnetic pressure supported disks \citep{ToutPringle1992,Pariev+2003,BegelmanPringle2007}, which can help solve many puzzles in classical thin disk models \citep{Begelman+2015,Sadowski2016,JIA19a,DexterBegelman2019}. Recent galaxy scale or cosmological zoom-in simulations also show that the accretion disks at the small scales can end up being dominated by magnetic fields even when initialized with weak magnetic fields 
on the galaxy scale \citep{Hopkins+2024,Guo+2024}. 

Once magnetic fields are introduced to control 
the vertical structures of accretion disks, it makes the predictions of all these models depend on the unknown structure and strength of 
these fields. One constraint on magnetic fields in accretion disks is from polarization \citep{BarnierDone2024}. Another powerful observational test is the spectrum emitted from the disk. 
Based on numerical simulations, it is generally true that stronger magnetic fields cause the inflow speed to be larger and thus result in smaller surface density for a given mass accretion rate. One extreme example is that the inflow speed gets close to the free-fall speed for the strongly magnetized disks shown in \cite{Hopkins+2024}.
For a typical $10^8$ solar mass black hole with Eddington mass accretion rate $2.3$ solar mass per year, if the ratio between the inflow speed and Keplerian speed is $f$, then the electron scattering optical depth across the disk is $1.76/f$ at 100 gravitational radii. The radiation spectrum emitted by accretion disks with such a low optical depth will likely be very different from the spectrum given by the optically thick standard thin disks. Therefore comparing observed AGN spectra with predictions of these accretion disk models will be a very powerful way to distinguish them. 

It is the goal of this paper to carry out a series of numerical experiments to test under what conditions the disk around supermassive black holes will be radiation or magnetic pressure supported for a wide range of mass accretion rates. These simulations with self-consistent thermodynamics as determined by the radiation transport calculations will also provide important clues on future physical models of AGN accretion disks, which can be very different from previous MHD simulations with an assumed equation of state. We also calculate the typical radiative spectrum from these different disks. Previous global radiation MHD simulations for AGNs typically cover the innermost region around the black holes \citep{JIA19a,JIA19b}, where hard X-rays are produced. Recently, frequency-dependent radiation MHD simulations have been attempted for X-ray binaries \citep{Fragile+2023,Roth+2025}. Since the extreme ultraviolet to soft X-ray component is estimated to be produced from a region $\approx 50-200$ gravitational radii from the black hole \citep{Jinetal2017}, which is also the radial range where thermal ultraviolet photons are expected to be produced, we have specifically designed our new simulation domain to extend out to this region of the disk. We neglect the innermost regions inside 50 gravitational radii (which we have simulated before in \citealt{JIA19a,JIA19b}) as including them would place a time step constraint that would make the simulations prohibitively expensive.

The remainder of this paper is organized as follows.
In Section \ref{sec:setup}, we describe the numerical setup for our 
simulations. The overall evolutions of these simulations are summarized in Section \ref{sec:overall}. In Section \ref{sec:vertical}, we describe the vertical structures of the disks in detail. Radiation spectra calculated for these disks are shown in Section \ref{sec:spectrum}. Finally, in Section \ref{sec:discussion}, we summarize the key results and discuss the implications. 






\begin{table*}[h]
	\caption{Simulation Summaries}
	\begin{center}
		\begin{tabular}{cccccccccc}
			\hline
			Runs & $\dot{M}/\     \dot{M}_{\rm Edd}$ & $r_i/\rg$ & $B$ Loop & $\rho_i/\rho_0$ & $P_i/P_0$ & $a_0$ & $<P_g/P_B>_{\rho}$ & $<P_r/P_B>_{\rho}$ & $\tau_i$\\
			\hline
			\twelve        &	  0.03	    & 	300  & \text{Single} & 0.2 & 9.6 & $10^{-4}$ & 3.25 & $2.16\times 10^2$ & $7.32\times 10^3$\\	
			\sixteen     	&	 $20 \sim 1$	    & 	 400	& \text{Single} & 1 & 108 & $4\times 10^{-4}$ & 0.15 & 11.87 & $1.26\times 10^5$ \\	
			\sixteenB        &	  0.6 &	400  & \text{Double}  & 1 & 108 & $4\times 10^{-4}$ & 3.55 & $2.91\times 10^2$ & $1.26 \times 10^5$\\	
			\eighteenB     	&	  $1-10$	    & 	400 	 & \text{Double}  & 10 & 1080 & $2\times 10^{-3}$ & 0.25 & 11.51 & $1.55\times 10^6$\\	
			\hline
        \end{tabular}
            
        \smallskip\footnotesize\centering
        The Eddington accretion rate in the second column is here defined with a ten percent radiative efficiency: $\dot{M}_\mathrm{Edd}\equiv L_\mathrm{Edd}/(0.1 c^2)$. \\ The parameters in all the other columns are for the initial torus, as described in Section \ref{sec:setup}.
	\end{center}
	\label{Table:parameters}
\end{table*}




\section{Simulation Setup}
\label{sec:setup}
All the simulations presented in this paper are run on a spherical polar coordinate $(r,\theta,\phi)$ grid centered on a $10^8\msun$ black hole, using the public available radiation MHD code \texttt{Athena++} \citep{Stone2020} with the radiation transport module developed by \cite{Jiang+2014} and \cite{Jiang2021}. The full set of ideal MHD equations are solved, which are coupled to the radiation transport equations by evolving specific intensities over discrete angles. The simulations start with a rotating torus whose construction follows the same procedure as described in \cite{JIA19b}.  There are three adjustable parameters:  the maximum density $\rho_i$, the maximum thermal pressure $P_i$, and the equatorial radius of this density/thermal pressure maximum $r_i$.  The total thermal pressure $P_t$ of the initial torus is assumed to follow a barotropic 
relation $P_t\propto \rho^{4/3}$. Taking the radiation and gas to be in thermal equilibrium initially, the gas temperature is then computed from the requirement that the sum of gas pressure and radiation pressure be equal to $P_t$. The magnetic fields are initialized with a vector potential that only has a non-zero component along the azimuthal direction $A_{\phi}$. We consider two possible initial magnetic configurations (e.g. \citealt{Beckwith+2008}). The first is a single magnetic field loop, for which we set $A_{\phi}=a_0 P_t^{1/2}r^2\sin\theta$, where $a_0$ is an adjustable parameter, for the region $130\rg<r<1000\rg$ and $88^{\circ} < \theta < 92^{\circ}$. We also consider double magnetic field loops, where we change the vector potential to be $A_{\phi}=a_0 P_t^{1/2}r^2\cos\theta$. The net poloidal magnetic field across the disk midplane at each radius is non-zero for the single loop case and it is zero for the setup with double magnetic field loops.
The evolution of the torus is controlled by the ratio between magnetic pressure and thermal pressure of the accreted gas, the magnetic topology, and the total optical depth, which determines the cooling time. If gas from the initial torus is able to cool significantly before it is accreted, the resulting disk can end up being more magnetic pressure-dominated than the case where the cooling is inefficient. Notice that gas from the initial torus can flow towards the black hole much more quickly than what a steady state thin disk model would predict if the gas near the inner edge of the torus can rapidly lose its angular momentum due to different magnetic processes. Therefore the usual argument about the hierarchy of thermal and viscous time scales in thin disk models does not apply during the initial evolution. 

The parameters we adopt for the runs are summarized in Table \ref{Table:parameters}, where the averaged gas pressure $P_g$, magnetic pressure $P_B$ and radiation pressure $P_r$ are weighted by the mass in each cell, and $\tau_i$ is the total vertical Rosseland mean optical depth at the center of the torus $r_i$.
The inner radial boundary is at $50\rg$ and the outer radial boundary is located at $4.45\times 10^3 \rg$ for \sixteen, \sixteenB\ and \eighteenB. At the inner/outer boundaries, we copy all the quantities from the last active zones to the ghost zones when the gas and photons are moving inward/outward. This is designed to make sure the properties of the disks are continuous across the inner boundary, which also means the stress at the inner boundary is not necessarily zero. We do not allow any gas or photons to come into the simulation domain via the radial boundaries.
The polar and azimuthal directions $(\theta, \phi)\in[0,\pi]\times [0,2\pi]$ are uniformly resolved with 32 and 64 cells, respectively, at the root level for these three runs. Three levels of refinement were used to cover the region $r,\theta,\phi\in[50\rg,500\rg]\times [86.5^{\circ},93.5^{\circ}]\times [0,2\pi]$, reaching a resolution of $\Delta r/r=\Delta \theta=\Delta \phi=1.23\%$. The choice is made so that we have enough resolution to resolve the dynamics within $\approx 7^{\circ}$ from the midplane of the disk, where most of the accretion happens.  And in addition, the time step of the simulations will not be limited by small cell sizes near the polar region. We adopted a finite wedge in the azimuthal direction and higher resolution for the run \twelve\ in order to resolve the smaller disk scale height due to the smaller accretion rate. At the root level, $32$, $32$ and $4$ cells are used to cover the simulation domain $(r,\theta,\phi)\in[50\rg, 10^3\rg]\times [0,\pi]\times[0,0.39]$. Then the region $(r,\theta,\phi)\in[50\rg, 250\rg]\times [87^{\circ},93^{\circ}]\times[0,0.39]$ had 5 levels of refinement, reaching a resolution of $\Delta r=\Delta \theta=\Delta \phi=0.3\%$.

\begin{figure*}[htp]
	\centering
	\includegraphics[width=0.45\hsize]{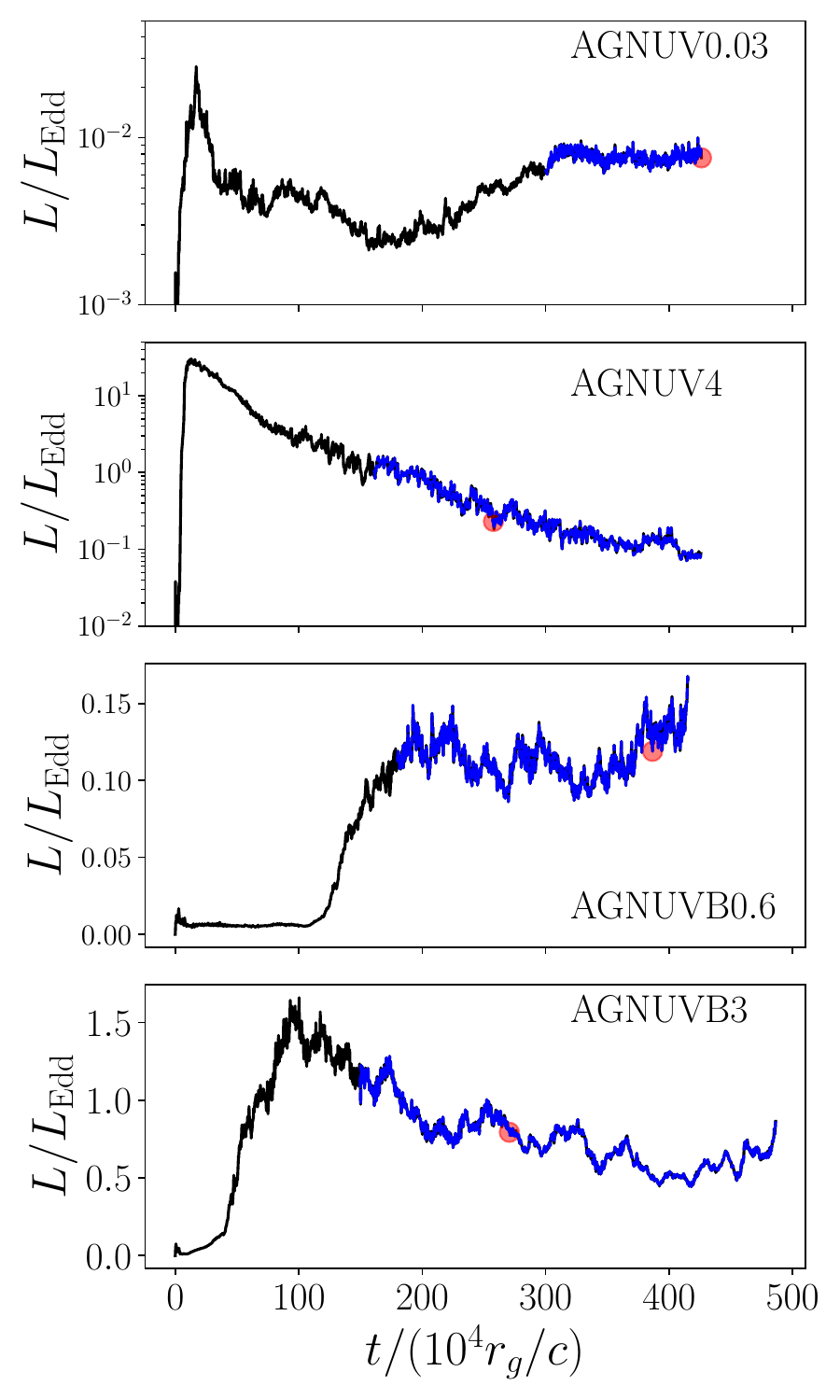}
	\includegraphics[width=0.45\hsize]{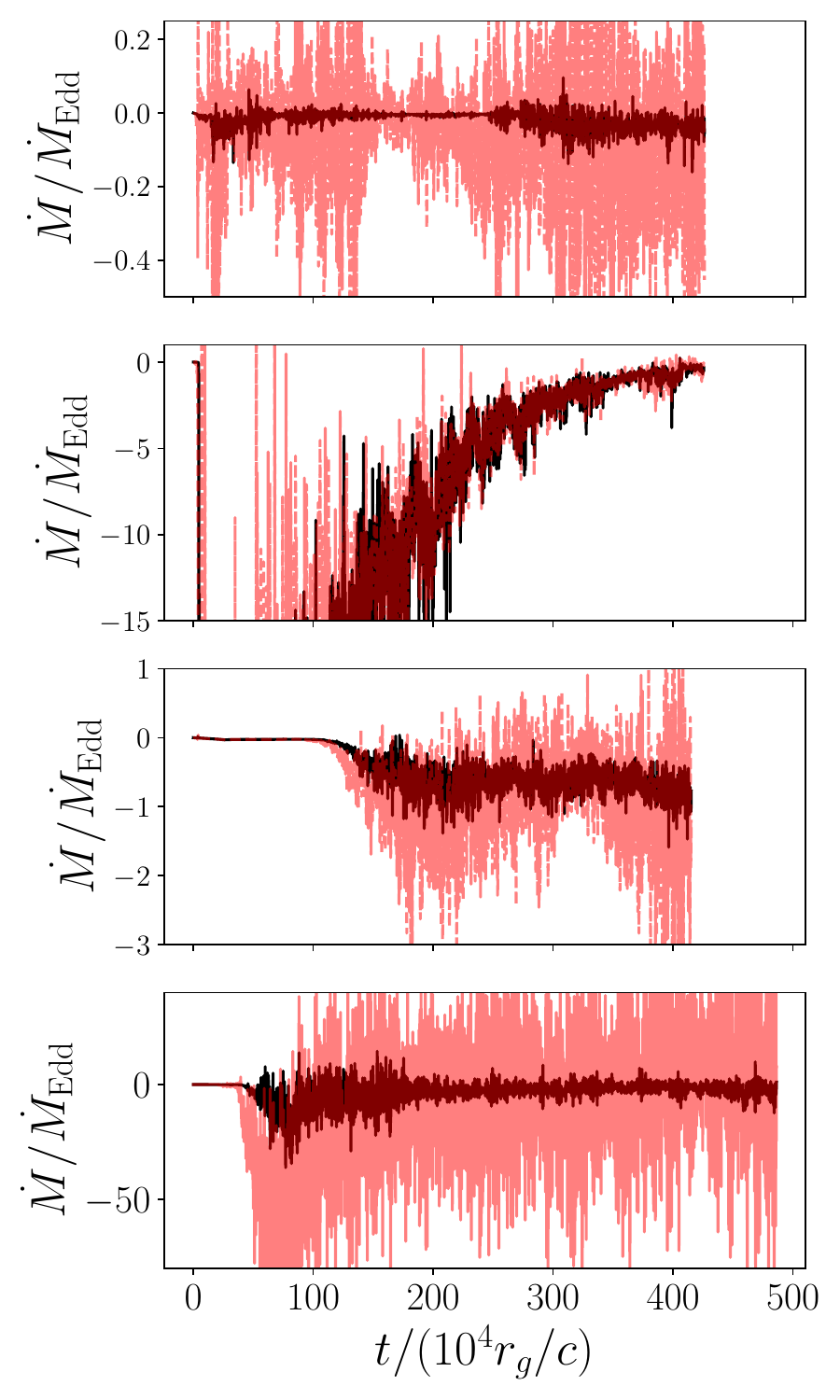}
	\caption{\emph{Left}: Histories of total luminosity emitted from the inner disks $(r<200\rg)$ of the four runs \twelve, \sixteen, \sixteenB\ and \eighteenB. Blue lines indicate the temporal intervals we used to calculate the time averaged properties. The snapshots we used to calculate the spectra are indicated by the red circles. \emph{Right}: Histories of mass accretion rate at radius $r=60\rg$ (black lines) and $r=150\rg$ (semi-transparent red lines) of the four runs \twelve, \sixteen, \sixteenB\ and \eighteenB\ (from top to bottom).  (Negative values of accretion rate correspond to infall.)}
	\label{lum_mdot_history}
\end{figure*}

We adopt the following fiducial units for all the simulations: $\rho_0=5\times 10^{-10}$ g/cm$^3$ for density, $T_0=10^5$ K for temperature, $P_0=6.93\times 10^3$ dyne/cm$^2$ for gas pressure and  $\rg=1.48\times 10^{13}$ cm for length. Radiation energy is often scaled by the fiducial number $a_rT_0^4=7.57\times 10^5$ erg/cm$^3$.

\section{Overall Evolution}
\label{sec:overall}

As MRI turbulence develops in the initial tori, matter and magnetic field in the inner regions accrete inward toward the inner boundary at $r=50r_\mathrm{g}$, eventually establishing an approximate steady-state.  The different physical parameters in the initial tori cause major differences in the final steady state flows.  During the early evolution of the two simulations \twelve\ and \sixteenB, the inflow time into the inner regions exceeds the cooling time, and these simulations lose thermal pressure support and become magnetically dominated before a steady state is established.  In simulation \sixteen, the cooling time in the inner regions exceeds the inflow time initially, allowing the flow to maintain thermal pressure support for a short time.  However, as this simulation proceeds, rapid accretion due to a large scale magnetocentrifugal wind depletes the surface mass density in the inner regions, reducing the cooling time and causing a loss of thermal pressure support.  A magnetically dominated region then develops and spreads outward in radius as time goes on. Simulation \eighteenB\ starts out with a much higher initial density and thermal pressure than the other three simulations, and its cooling time exceeds the inflow time into the inner regions, allowing it to retain radiation pressure support as it reaches a steady state.

\subsection{Accretion Rates and Luminosities}

Histories of luminosity output from the region inside $200\rg$ of the four runs are shown in the left panels of Figure \ref{lum_mdot_history}, while the evolution of net mass accretion rates at $60\rg$ and $150\rg$ are shown in the right panels of Figure \ref{lum_mdot_history}. 

All four simulations, to varying degrees, exhibit outward energy flows of Poynting flux, enthalpy flux, and mechanical energy flux, but radiative luminosity always dominates the overall energetics.  Indeed, all four simulations are radiatively efficient, in the following sense.  Standard nonrelativistic, stationary geometrically thin accretion disk theory would predict a radiative efficiency for luminosity emitted between  $r_{\rm in}$ and $r_{\rm out}$ to be
\begin{eqnarray}
    \eta&\equiv&\frac{L}{\dot{M}c^2}=\frac{1}{\dot{M}c^2}\int_{r_{\rm in}}^{r_{\rm out}}dr2\pi r\frac{3GM\dot{M}}{4\pi r^3}\Biggl[1-\left({\frac{r_{\rm in}}{r}}\right)^{1/2}\cr
    &+&{\frac{2\pi r_{\rm in}^2W_{r\phi,{\rm in}}}{\dot{M}(GMr)^{1/2}}\Biggr]}\cr
    &=&\frac{r_{\rm g}}{2r_{\rm in}}\left[1-\frac{3r_{\rm in}}{r_{\rm out}}+2\left(\frac{r_{\rm in}}{r_{\rm out}}\right)^{3/2}\right]\cr
    &+&\frac{2\pi (GMr_{\rm in})^{1/2}W_{r\phi,{\rm in}}}{\dot{M}c^2}\left[1-\left(\frac{r_{\rm in}}{r_{\rm out}}\right)^{3/2}\right],
    \label{eq:etathindisktheory}
\end{eqnarray}
where $W_{r\phi,{\rm in}}$ is the vertically integrated stress at $r_{\rm in}$ and $\dot{M}$ is the net mass accretion rate through the disk.  If we neglect this for the moment, then with our inner boundary of $r_{\rm in}=50r_{\rm g}$, this gives 0.5\% for $r_{\rm out}=200r_{\rm g}$.  We compute the ratio of instantaneous luminosity leaving $r_\mathrm{out}=200r_\mathrm{g}$ to the mass accretion rate through the inner boundary $r_\mathrm{in}=50r_\mathrm{g}$ from the simulation data, and find that equation (\ref{eq:etathindisktheory}) matches the data well ($\eta=0.5$\%)  for long time periods in the single-loop simulation \sixteen.  This is true even though both the luminosity and accretion rate decline by roughly two orders of magnitude over the course of the simulation as the initial mass reservoir is drained.

\begin{figure}[htp]
	\centering
	\includegraphics[width=1.0\hsize]{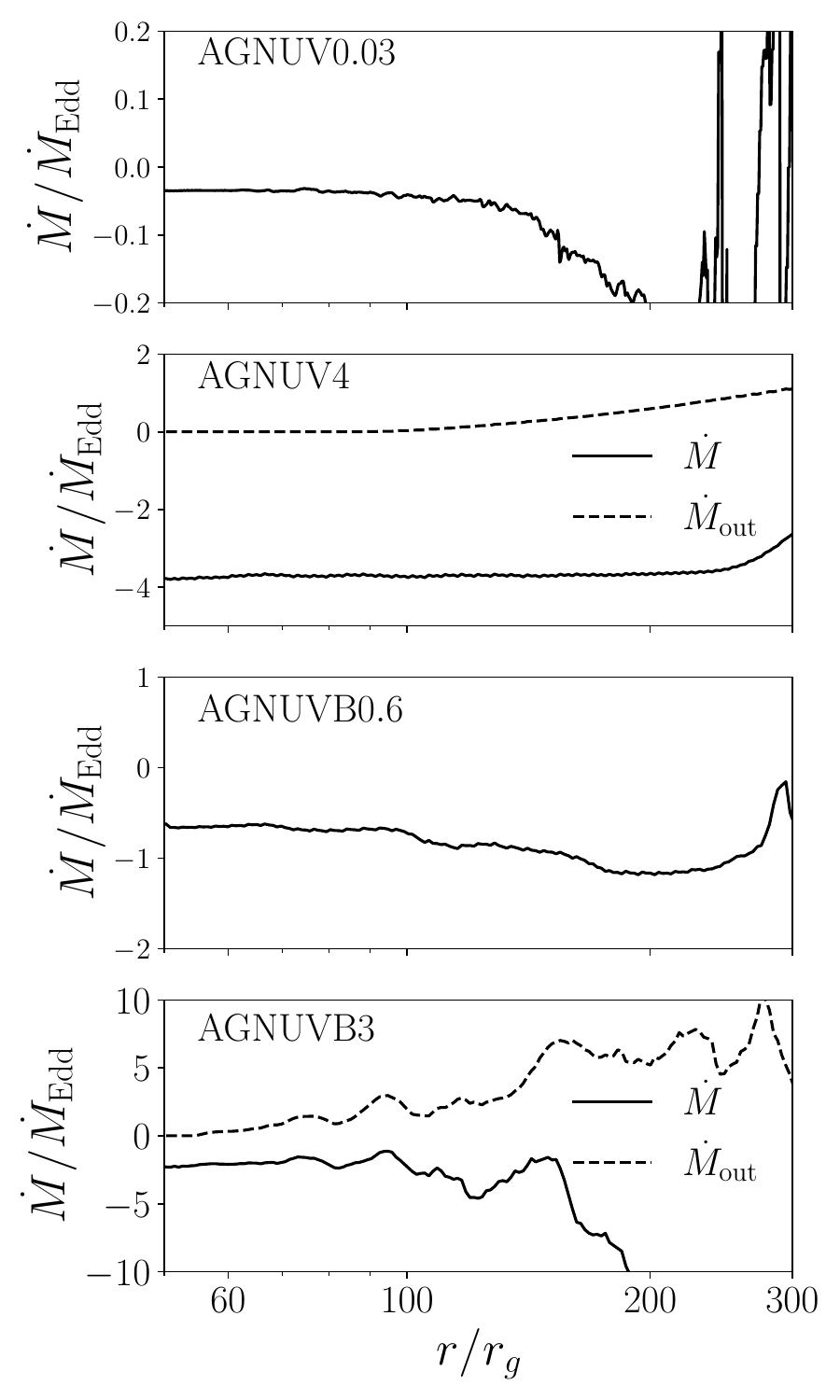}
	\caption{Radial profiles of time-averaged net mass accretion rates (solid black lines) for the four runs \twelve, \sixteen, \sixteenB\ and \eighteenB\ (from top to bottom panels). The dashed black lines in the second and fourth panels are the mass fluxes moving out, which are basically 0 for other two runs. The time averages are done for the time intervals $[3.0,4.3]\times 10^6\ \rg/c$,  $[1.6,4.3]\times 10^6\ \rg/c$, $[1.8,4.2]\times 10^6\ \rg/c$, $[1.5,4.9]\times 10^6\ \rg/c$ of the four runs, respectively.}
	\label{mdot_radial}
\end{figure}

Figure~\ref{mdot_radial} shows time-averaged radial accretion rate profiles for all four simulations. The temporal windows we use to perform the time average are indicated by the blue lines in the left panel of Figure \ref{lum_mdot_history}.
Inflow equilibrium, in the sense that the accretion rate is independent of radius (though not time), is in fact established inside $r=200r_\mathrm{g}$ for run \sixteen, which is why we chose that radius for $r_{\rm out}$ in our efficiency calculation.  At late times, the first term (no inner stress) also matches the measured efficiency of run \twelve.  However,
runs \sixteenB\/ and \eighteenB\/ have somewhat higher efficiencies, likely because of significant contributions from stress at the inner boundary.

\begin{figure*}[htp]
	\centering
	\includegraphics[width=0.45\hsize]{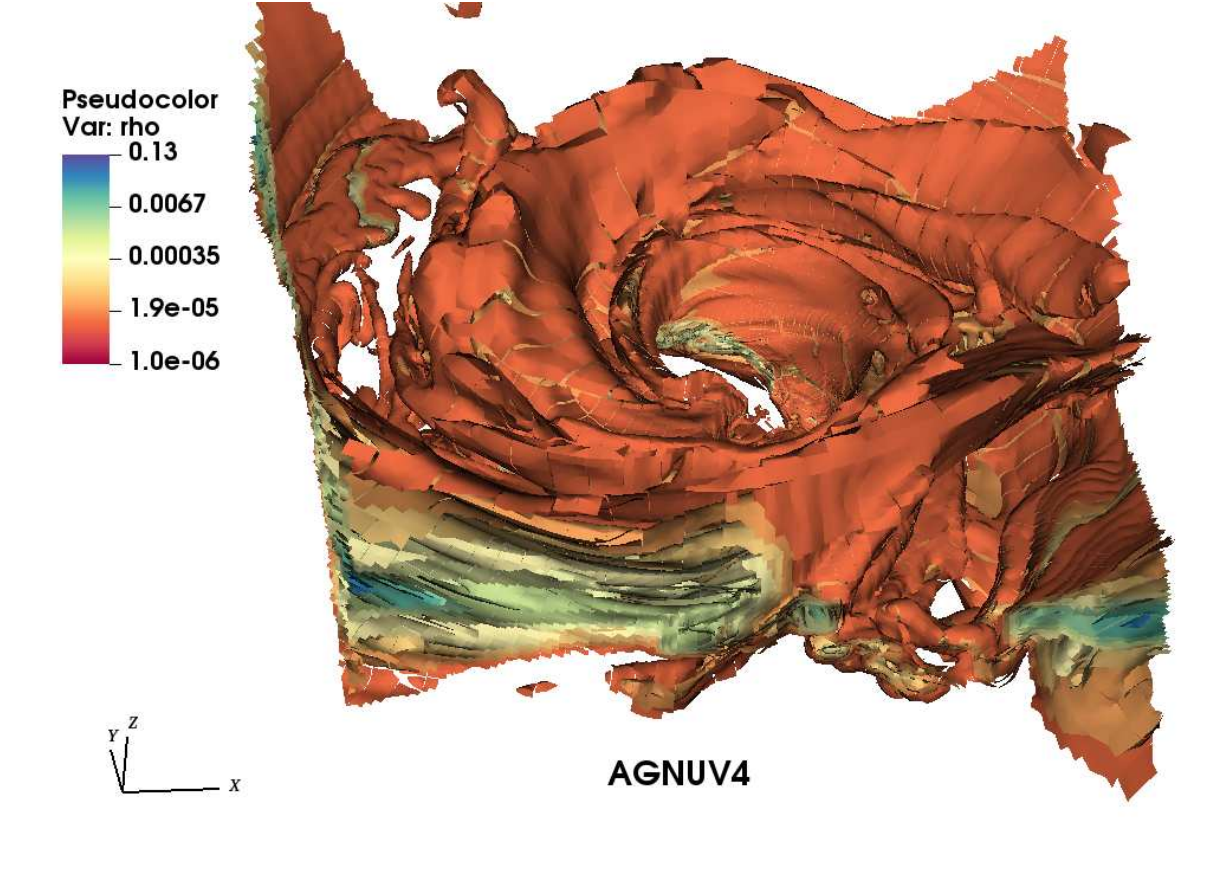}
	\includegraphics[width=0.45\hsize]{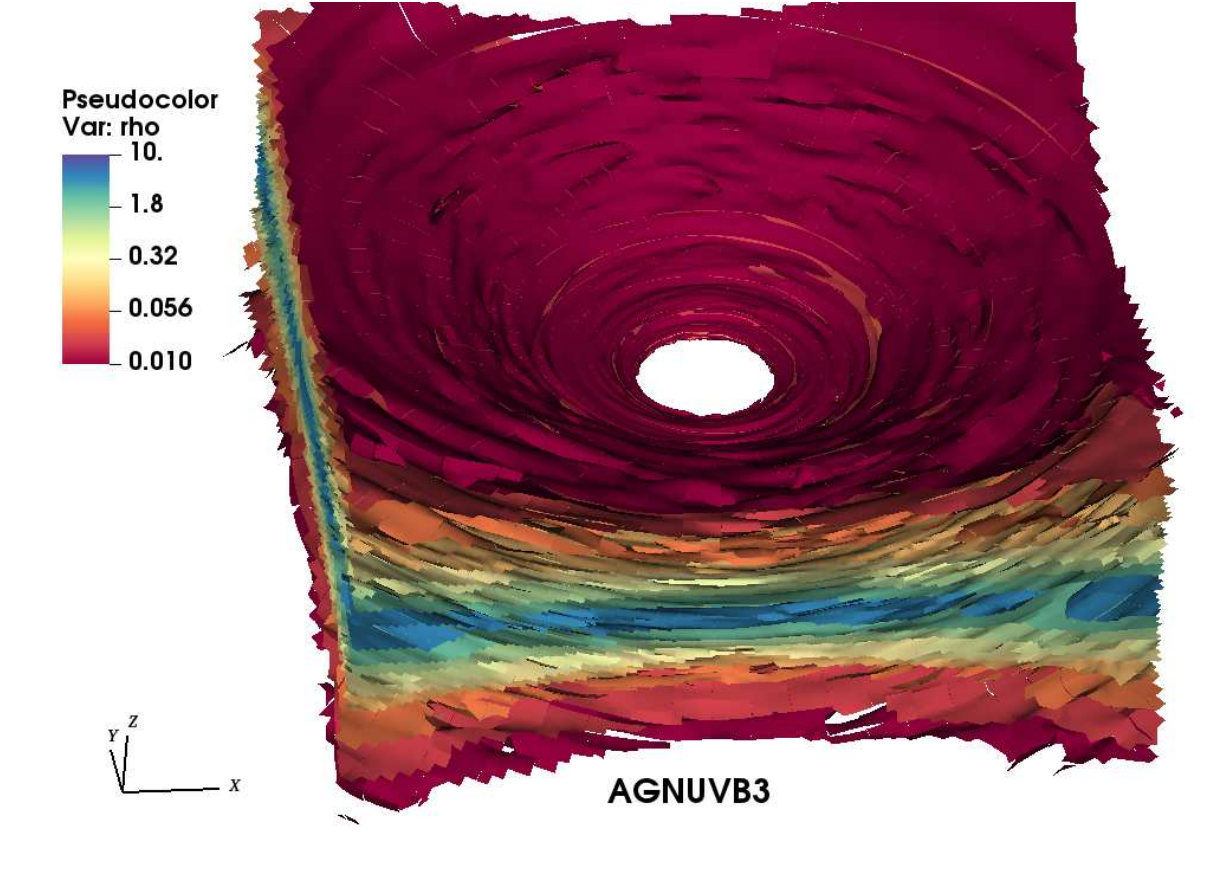}
	\caption{Three dimensional structures of density 
    for a snapshot at time $2.58\times 10^6\ \rg/c$ of the run \sixteen\ (left) and a snapshot at time $2.71\times 10^6\ \rg/c$ of the run \eighteenB\ (right). Density is shown as iso-surfaces within $300\rg$.}
	\label{AGNWedge_16_18B_3D}
\end{figure*}

\subsection{Spatial Structure}

The three-dimensional (3D) structure of the disk is quite different across all four simulations.  Iso-surfaces of density in 3D for snapshots of the single-loop run \sixteen\ and the double-loop run \eighteenB\ are shown in Figure \ref{AGNWedge_16_18B_3D}. Azimuthal variations of density are relatively small for the run \eighteenB\ and it shows a typical rotationally supported disk.
However, the disk in the run \sixteen\ shows a much larger azimuthal variation with significant spiral like structures. 

\begin{figure*}[htp]
	\centering
	\includegraphics[width=1\hsize]{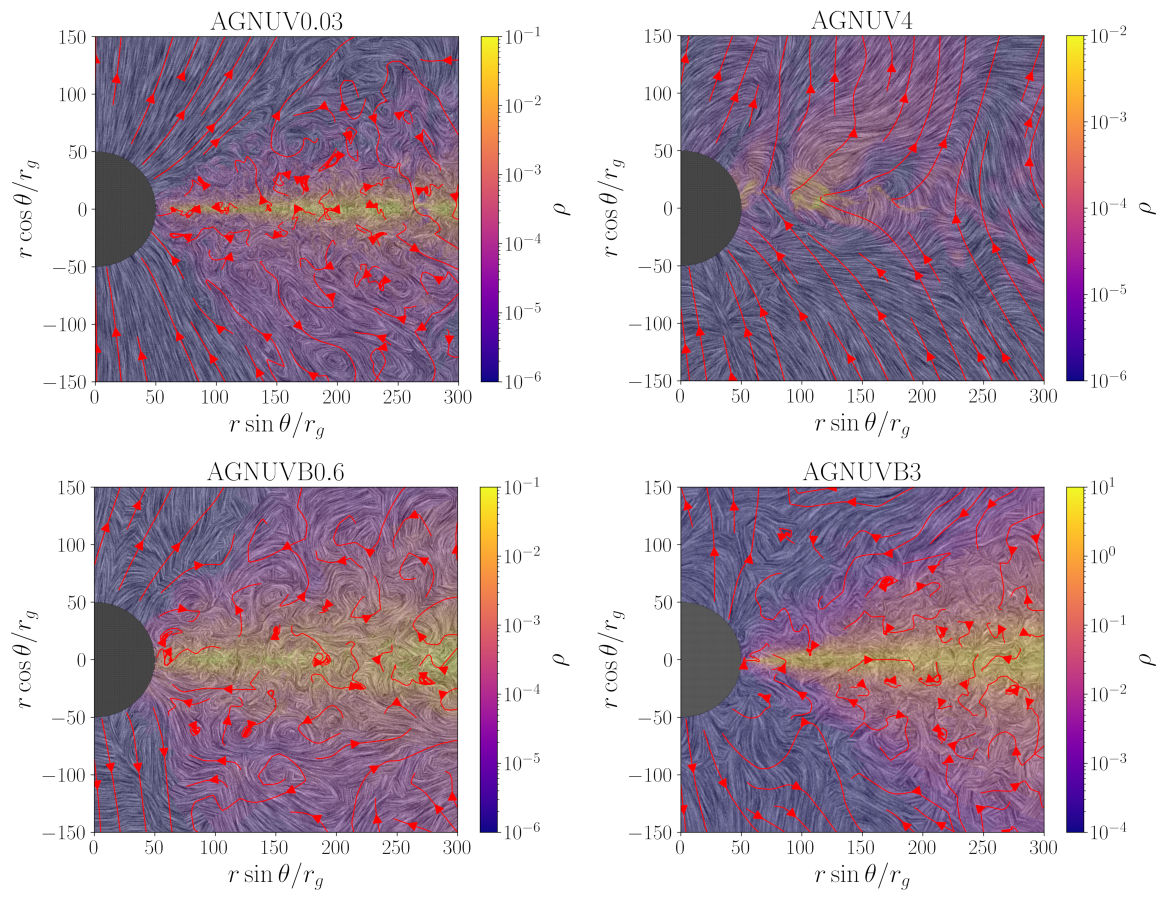}
	\caption{Vertical slices of density for snapshots at time $4.26\times 10^6\ \rg/c$, $2.58\times 10^6\ \rg/c$, $3.87\times 10^6\ \rg/c$ and $2.71\times 10^6\ \rg/c$ respectively for the four runs. The over-plotted gray lines are the line integral convolution of the velocity, which is an efficient way to show the velocity streamlines in high spatial resolution. The red lines represent streamlines of magnetic fields.}
	\label{snapshot_rho_B_slice}
\end{figure*}


Representative snapshots of the poloidal density and magnetic field lines in each simulation are shown in Figure \ref{snapshot_rho_B_slice}.  (The times of these snapshots are indicated by the filled red circles in the left panel of Figure~\ref{lum_mdot_history}.)  Magnetic fields accumulated near the rotational axis roughly follow the topology of the initial magnetic fields in the torus. Except for the run \sixteen, all the other three runs have a rotationally supported disk formed near the midplane. Comparing the two magnetically supported disks \twelve\ and \sixteenB, the density scale height is considerably smaller in run \twelve\ due to its lower accretion rate. Density is also strongly peaked near the midplane for the run \eighteenB\ despite its larger mass accretion rate. As we discuss below, this is because the disk in this run is supported by radiation pressure. 


Mean Maxwell stresses dominate over turbulent Maxwell plus Reynolds stresses in the magnetically dominated simulations \twelve\ and \sixteenB.  Nevertheless, snapshots of the poloidal field line structures in these two simulations in Figure~\ref{snapshot_rho_B_slice} show significant spatial disorder, indicative of turbulence.  The toroidal field is much more ordered, reflecting the fact that the time-averaged poloidal field in these two simulations is more coherent.  The polarity of the time-averaged poloidal field in \sixteenB\ is consistent with inward advection of the initial double-loop field topology, but there is strong evidence in the initial evolution of some sort of dynamo amplification of the field to levels well above the result of simple flux-frozen advection.


\begin{figure*}[htp]
	\centering
	\includegraphics[width=0.45\hsize]{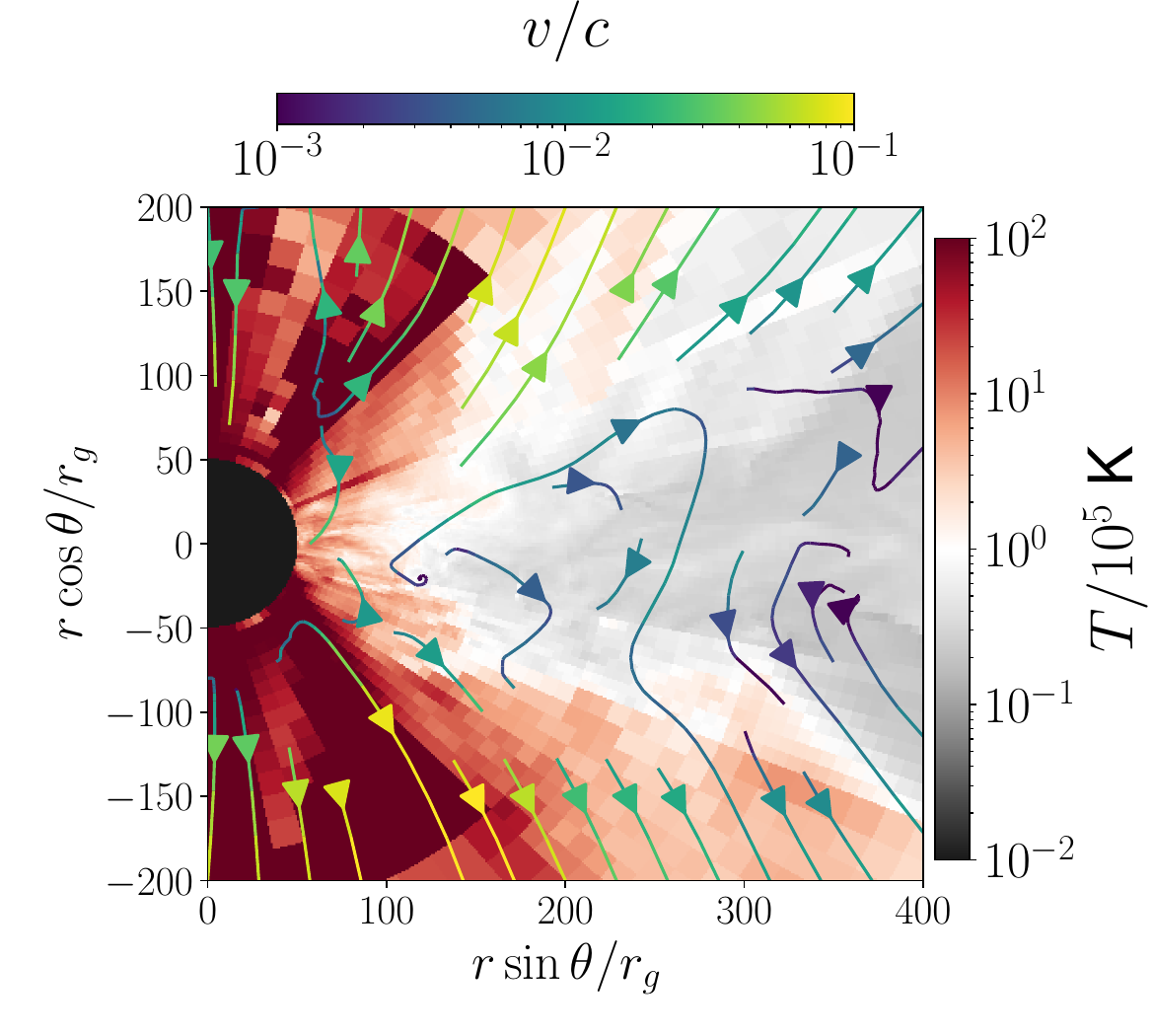}
	\includegraphics[width=0.45\hsize]{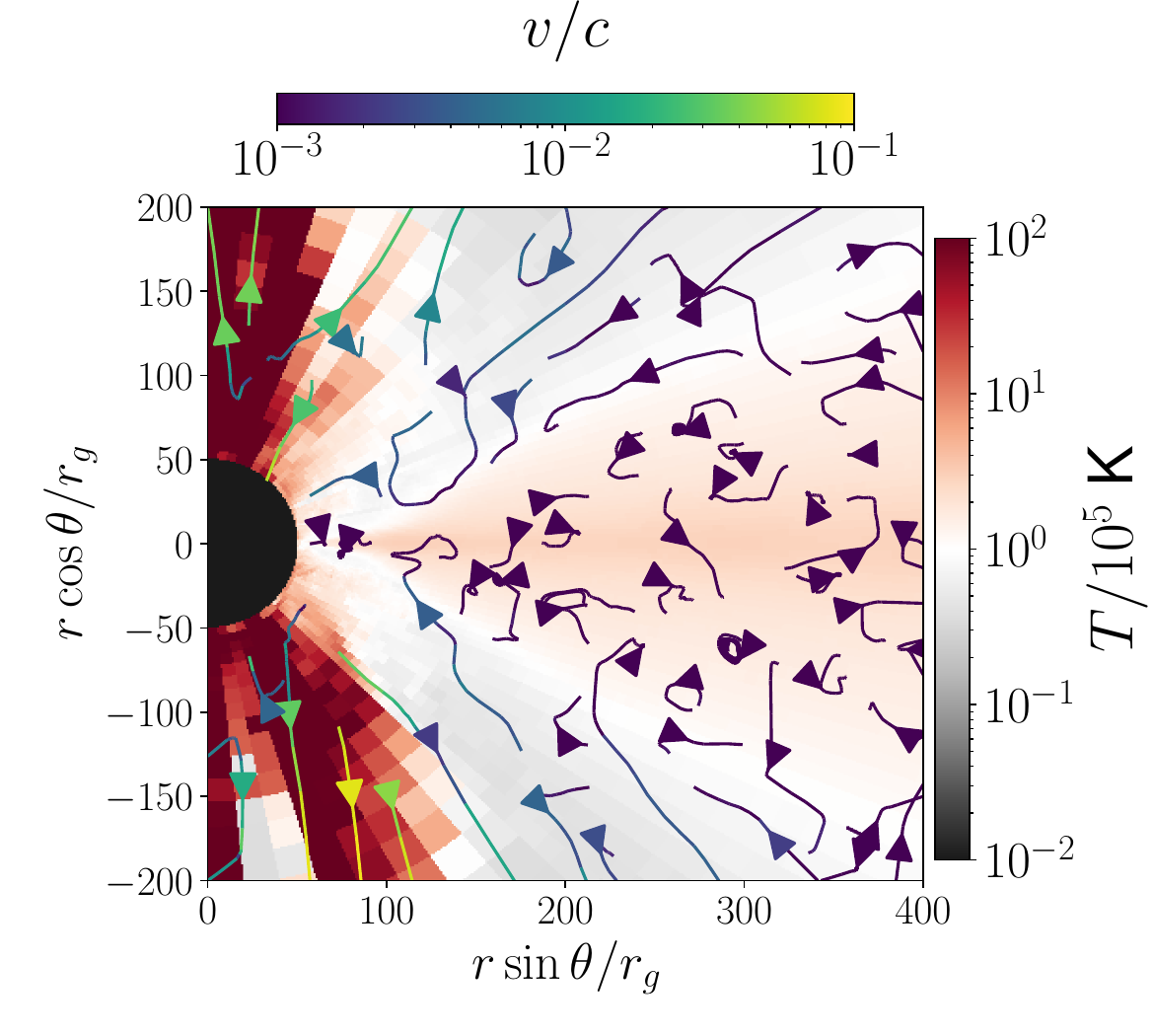}
	\caption{Azimuthally averaged gas temperature for a snapshot at time $2.58\times 10^6\ \rg/c$ of the run \sixteen \ (left) and a snapshot at time $2.71\times 10^6\ \rg/c$ of the run \eighteenB \ (right). The streamlines show the azimuthally averaged flow velocities (the $v_r$ and $v_{\theta}$ components).}
	\label{wedge18B_16_Tgas}
\end{figure*}

In contrast, the poloidal field lines in the magnetically dominated regions of the single-loop run \sixteen\ (upper right panel of Figure~\ref{snapshot_rho_B_slice}) are highly ordered.  The polarity of the radial field component flips across the midplane, and shearing of this field also causes the toroidal field to flip polarity across the midplane.  This polarity appears to reflect the stretching of the initial single-loop magnetic field inward by the accretion flow.  Like the magnetically dominated simulation \sixteenB, run \sixteen\ has significant mean field $B_rB_\phi$ stresses transporting angular momentum radially outward.  In addition, though, there is substantial outward vertical transport of angular momentum in a magnetocentrifugal wind.  This is not present in the double-loop simulation \sixteenB.  It is these extra mean field magnetic stresses that drive the high early accretion rate in run \sixteen, which decreases in magnitude with time as it drains the torus reservoir.  The large-scale ordered magnetic field is also likely what is preventing a disk-like density distribution from forming at the midplane.

Run \eighteenB, with its much higher initial density and thermal pressure, persists in being radiation pressure dominated, with midplane plasma $\beta$'s ranging between ten and several hundred.  Like \sixteenB, it started with a double-loop topology in the initial torus, but the radial and azimuthal magnetic field flip polarity back and forth in time in agreement with the classic weak-field MRI butterfly dynamo \citep{BRA95}.  Turbulent, not mean-field, Maxwell stresses dominate the outward angular momentum transport.  The same behavior occurs in the thermal pressure dominated 
epochs of run \sixteen.


Gas temperature and velocity streamlines for the same snapshots of the two runs \sixteen\ and \eighteenB\ are shown in Figure \ref{wedge18B_16_Tgas}. Near the midplane, gas temperature varies around $\approx 1-3\times 10^5$ K and it increases with increasing mass accretion rate. This is quite consistent with the value we expect if the locally dissipated accretion power is thermalized. 
The magneto-centrifugal wind in run \sixteen\ is evident in the velocity streamlines in the left panel of Figure~\ref{wedge18B_16_Tgas}.  
This wind shows up for a wide radial range with typical outflow speeds varying between $\approx 0.01c$ and $0.1c$. The average mass flux carried by the wind in \sixteen\ is negligible compared to the mass accretion rate for $r<150\rg$, as shown in the second panel of Figure \ref{mdot_radial}. In the radiation pressure-dominated run \eighteenB, there is also an outflow, but it is driven by the strong radiation force inside $\approx 100\rg$.

\begin{figure*}[htp]
	\centering
	\includegraphics[width=0.45\hsize]{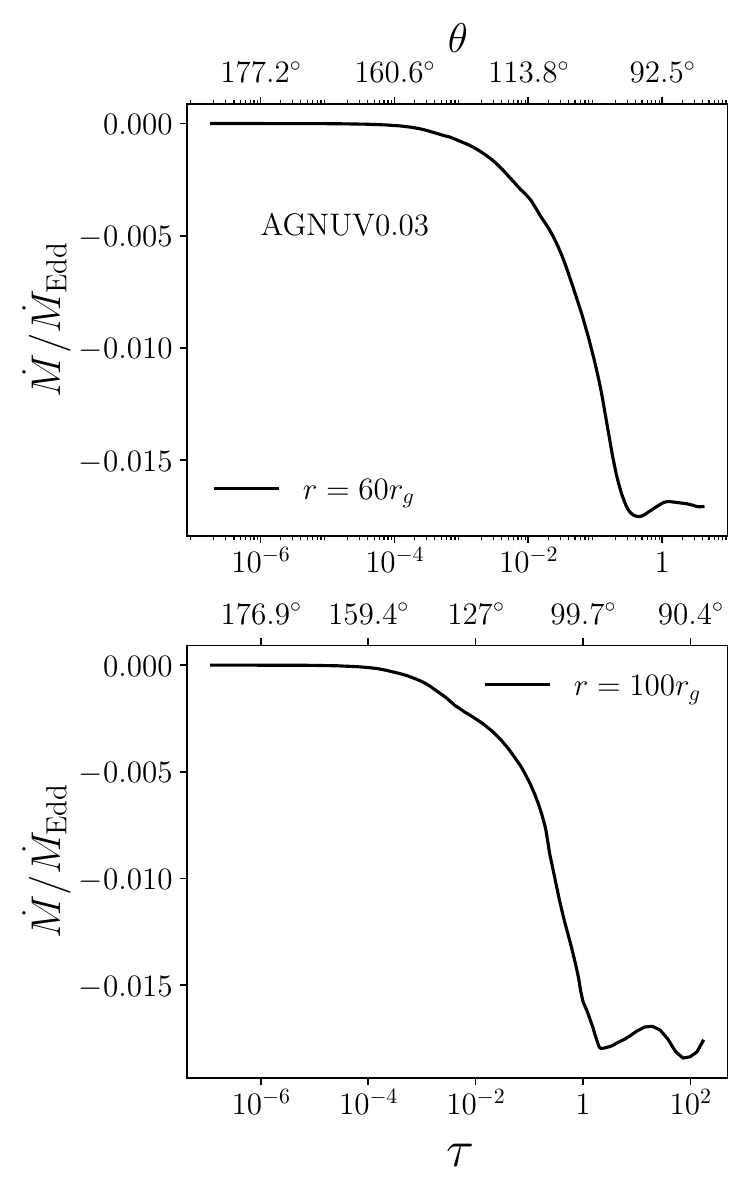}
	\includegraphics[width=0.45\hsize]{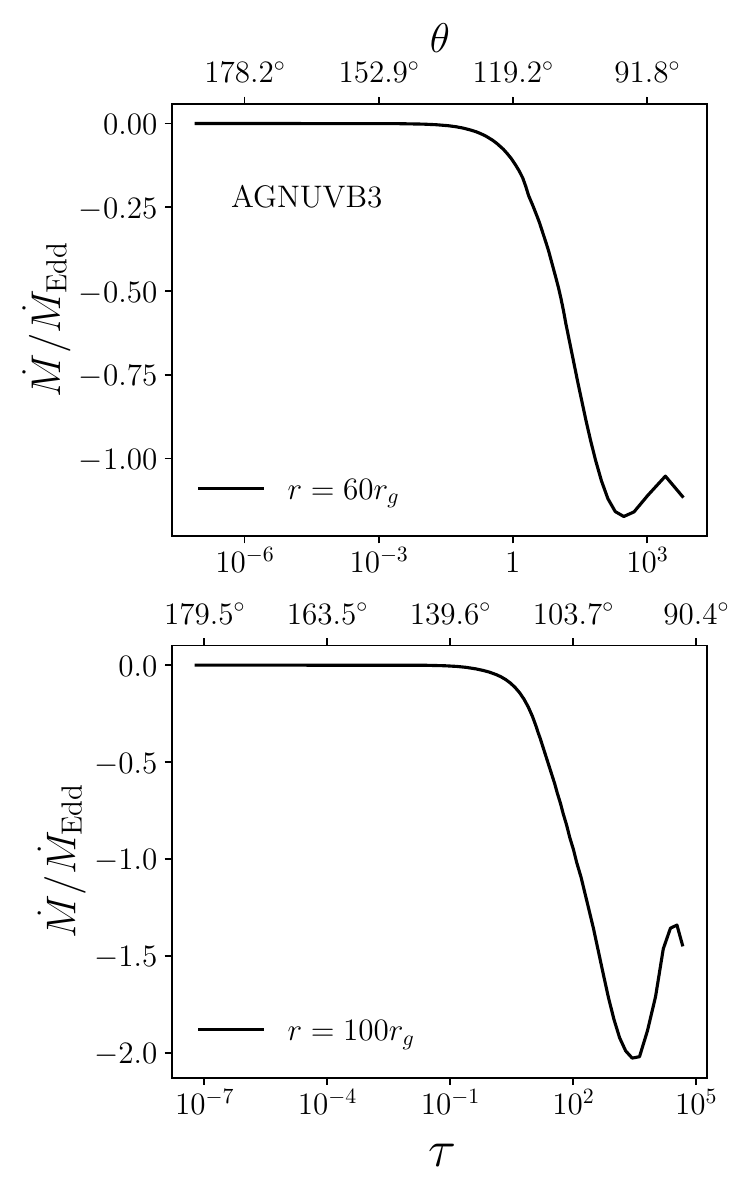}
	\caption{Time averaged mass accretion rates integrated between  $\theta=180^{\circ}$ and each polar position $\theta$ at radius $r=60r_\mathrm{g}$ (top row) and $100r_\mathrm{g}$ (bottom row) for the simulations \twelve\ (left) and \eighteenB\ (right).  Corresponding Rosseland mean optical depths are shown on the bottom axes of each panel. The fluctuations of $\dot{M}$ near the midplane are caused by the inflow and outflow motions at the radii indicated.}
	\label{mdot_tau}
\end{figure*}


\section{Vertical Structures of the Disks}
\label{sec:vertical}

We now examine the average vertical distribution of various 
physical quantities that are important in shaping the observable properties of these distinct simulated accretion flows.  For any quantity $a$, we first average it azimuthally, which is referred to as $\langle a \rangle$, and then we time-averaged to produce $\overline{\langle a\rangle}$.  The epochs used in the time averages are indicated by the blue lines shown in Figure \ref{lum_mdot_history}.  These time intervals are chosen to represent approximate steady states for each run, with the exception of the magneto-centrifugal wind run \sixteen\ which shows a continuous decline in mass and luminosity.

\begin{figure}[htp]
	\centering
	\includegraphics[width=1.0\hsize]{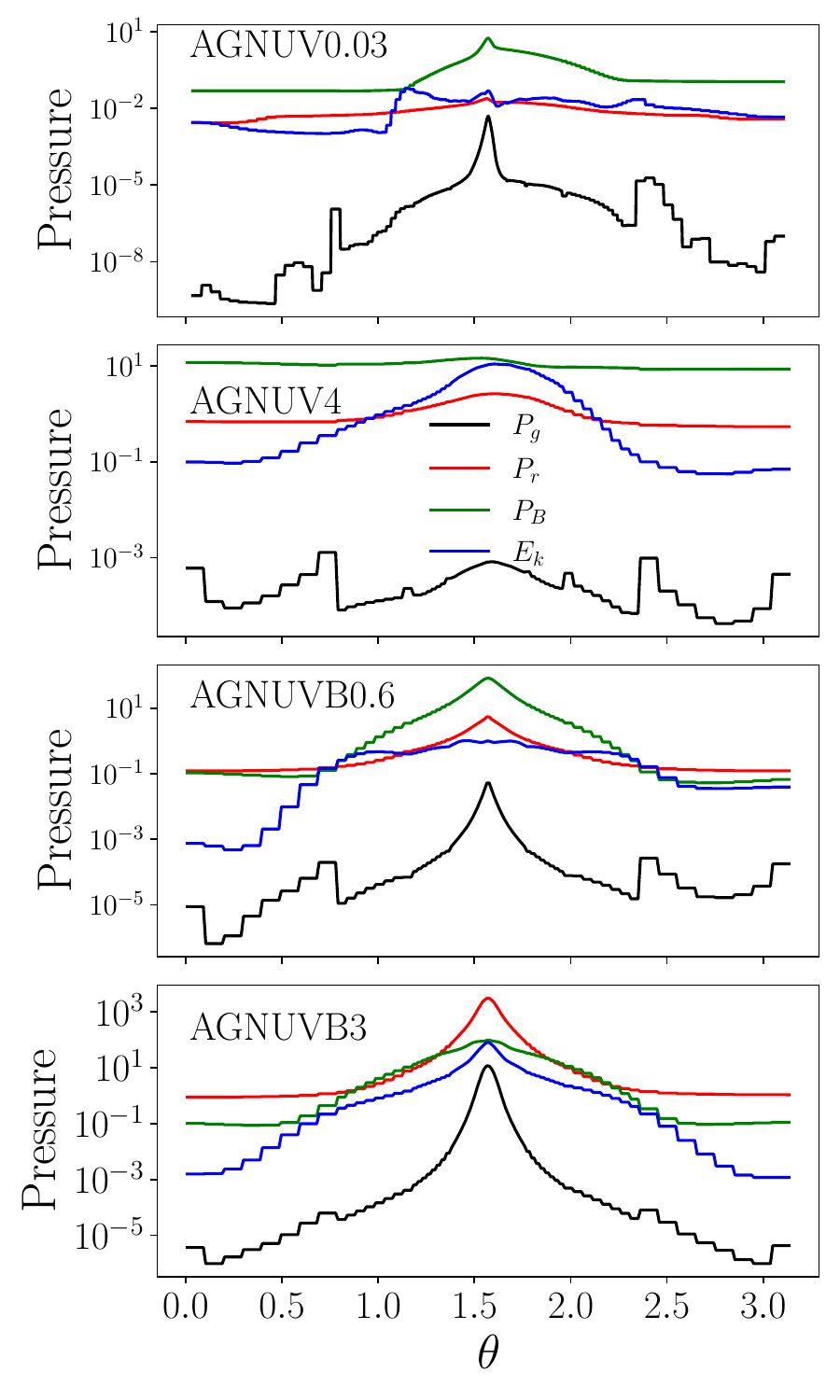}
	\caption{Time and azimuthally averaged vertical profiles of different pressure components at $r=100\rg$ for the four runs as labeled in each panel. The gas pressure (black), radiation pressure (red), magnetic pressure (green) and poloidal kinetic energy density (blue) are in units of $P_0=6.93\times 10^3$ dyne/cm$^2$. }
	\label{vertical_pressure}
\end{figure}

\begin{figure}[htp]
	\centering
	\includegraphics[width=1.0\hsize]{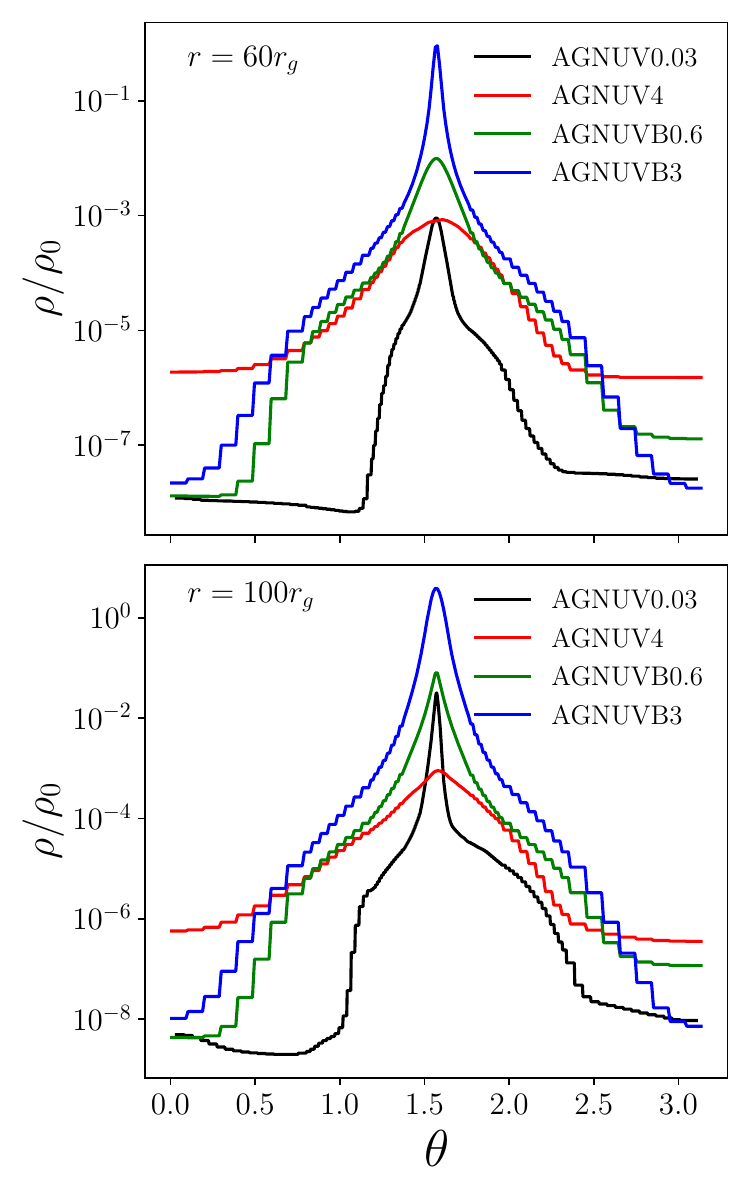}
	\caption{Time and azimuthally averaged vertical density profiles at radii $60\rg$ (top) and $100\rg$ (bottom) for the four runs. Densities are scaled with the fiducial density unit $\rho_0=5\times 10^{-10}$ g/cm$^3$.}
	\label{vertical_density}
\end{figure}

\begin{figure}[htp]
	\centering
	\includegraphics[width=1.0\hsize]{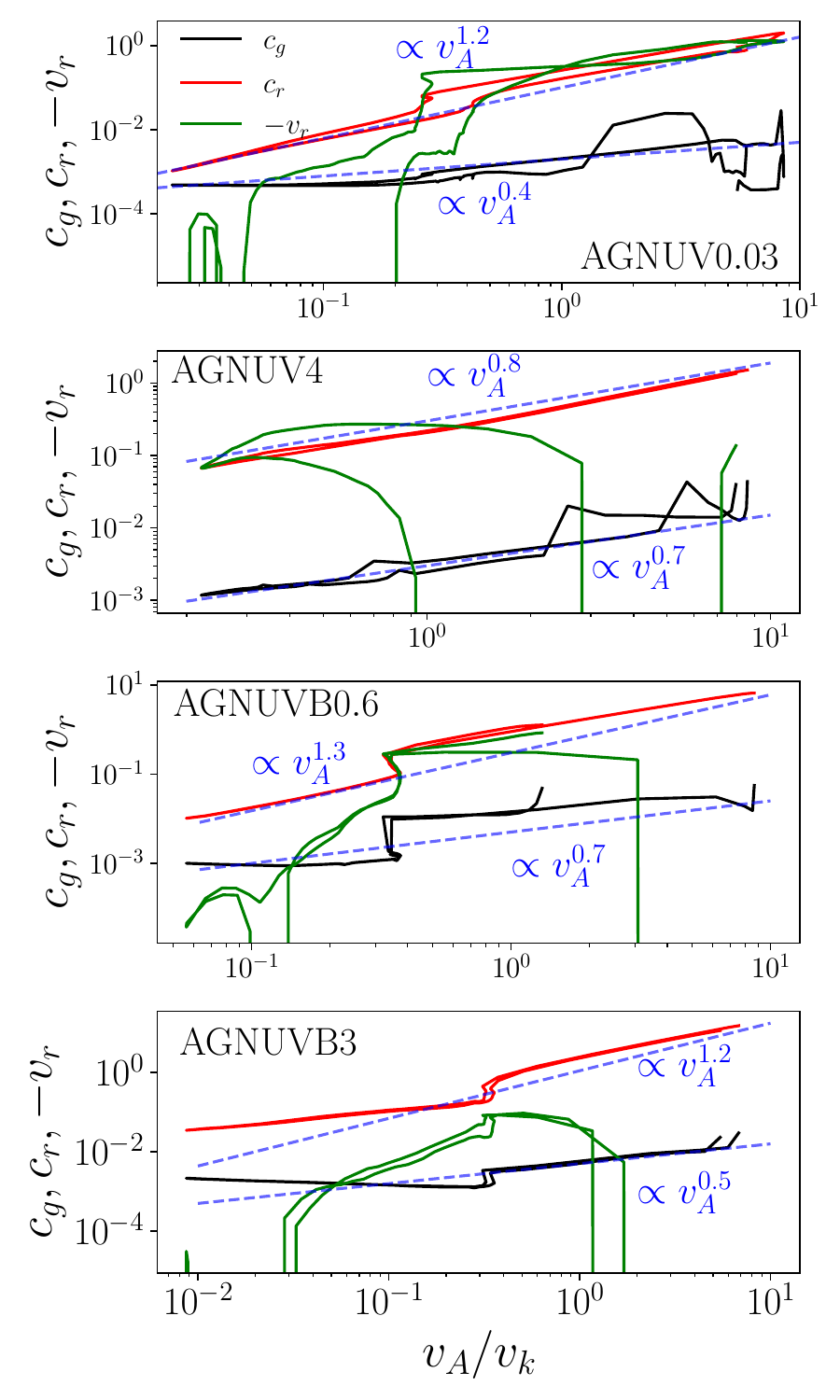}
	\caption{Azimuthally averaged vertical profiles of gas sound speed $c_g$ (black lines), radiation sound speed $c_r$ (red lines) and inflow speed $-v_r$ (green lines) as a function of Alfv\'en speed ($v_k$) at $r=100\rg$ for the four runs \twelve, \sixteen, \sixteenB\ and \eighteenB\ (from top to bottom). All the velocities are scaled with the Keplerian speed at the same radius. Each line has two segments that represent two sides of the disk from the midplane. The left side with the smallest $v_A$ corresponds to the disk midplane while the right side with the largest $v_A$ corresponds to the polar region. The dashed blue lines are power law scalings as indicated in each panel.}
	\label{velocity_va_relation}
\end{figure}

\subsection{Vertical Distributions of Mass Accretion Rate}
\label{sec:vertical_mdot}

We quantify the vertical distribution of mass accretion using the integrated mass flux
$\dot{M}(r,\theta)\equiv 2\pi r^2 \int_{\theta}^{\pi} \overline{\langle\rho v_r\rangle}(\chi) \sin\chi d\chi $.
Two examples at $r=60\rg$ and $100\rg$ for the two runs \twelve\ and \sixteenB\ are shown in Figure \ref{mdot_tau}. The integrated Rosseland mean optical depth from the south polar axis at each $\theta$ is shown at the bottom of each panel, while the corresponding polar angle $\theta$ is shown on the top horizontal axis. 
For the run \twelve\ with the lowest accretion rate, most of the accretion occurs in the region where the integrated optical depth from the pole is smaller than $1$, which happens around $5^{\circ}-10^{\circ}$ from the midplane for the two radii we chose. In contrast, for the run \sixteenB, the accretion happening in the optically thin region is negligible, and the location where $\tau=1$ is $\approx 30^{\circ}$ from the midplane. For this run, half of the accretion is located in the region $\approx 10^{\circ}-15^{\circ}$ from the midplane. Density scale heights are much smaller than these locations as we will demonstrate later. This is consistent with previous results that accretion occurs in a surface layer instead of the midplane for magnetic pressure-supported disks \citep{ZhuStone2018,ZhuJiangStone2019}. However, we demonstrate here in run \sixteenB\ that not all these ``surface accretion" flows are optically thin, which is necessary to increase the gas temperature due to local dissipation. Generally, the fraction of accretion in the optically thin region increases with lower surface density, corresponding to lower accretion rates. These trends are also consistent with previous local shearing box \citep{Jiangetal2014} or global \citep{JIA19a,Huangetal2023} radiation MHD simulations, even though they cover a different radial range of the disk. 

Both runs \sixteen\ and \eighteenB\ have similar polar distributions of mass accretion rate to that of \sixteenB, with the majority of accretion happening in the optically thick region. Half of the accretion is located within $10^{\circ}$ from the midplane for the run \sixteen\ while the corresponding number is $15^{\circ}$ for the run \eighteenB. The run \eighteenB\ also shows fluctuations of $\dot{M}$ near the midplane due to MRI turbulence, which does not happen for \sixteen. So it is generally true that more accretion can happen at the spatial location several density scale heights away from the disk midplane, which can happen for both magnetic pressure and radiation pressure supported disks.


Vertical profiles of the time and azimuthally averaged radiation pressure $\overline{\langle P_r\rangle }$, gas pressure $\overline{\langle P_g\rangle }$, magnetic pressure $\overline{\langle P_B\rangle }$ as well as the kinetic energy density $\overline{\langle E_k\rangle}$ at $r=100\rg$ for the four runs are shown in Figure \ref{vertical_pressure}. Here $\overline{\langle E_k\rangle }$ only includes the poloidal
velocities to avoid the large rotational velocity. Near the midplane, magnetic pressure is the dominant pressure in supporting the disks against vertical gravity for the three runs \twelve, \sixteen, and \sixteenB, while radiation pressure is larger than magnetic pressure by a factor of $\approx 100$ near the midplane in run \eighteenB. Gas pressure is always negligible for all the runs, as expected for this region of an accretion disk around a $10^8\msun$ black hole.

The kinetic energy densities associated with $v_r$ and $v_{\theta}$ are about a few percent of the dominant pressure component near the midplane for the three runs \twelve\ \sixteenB\ and \eighteenB, and are associated with the turbulent motions in the disk. The only exception is the run \sixteen, where $\overline{\langle E_k\rangle }$ is comparable to $\overline{\langle P_B\rangle }$ 
near the midplane. 
This is because the magneto-centrifugal wind in this run significantly enhances both the inflow and vertical velocities 
compared to the other runs. 

\begin{figure}[htp]
	\centering
	\includegraphics[width=1.0\hsize]{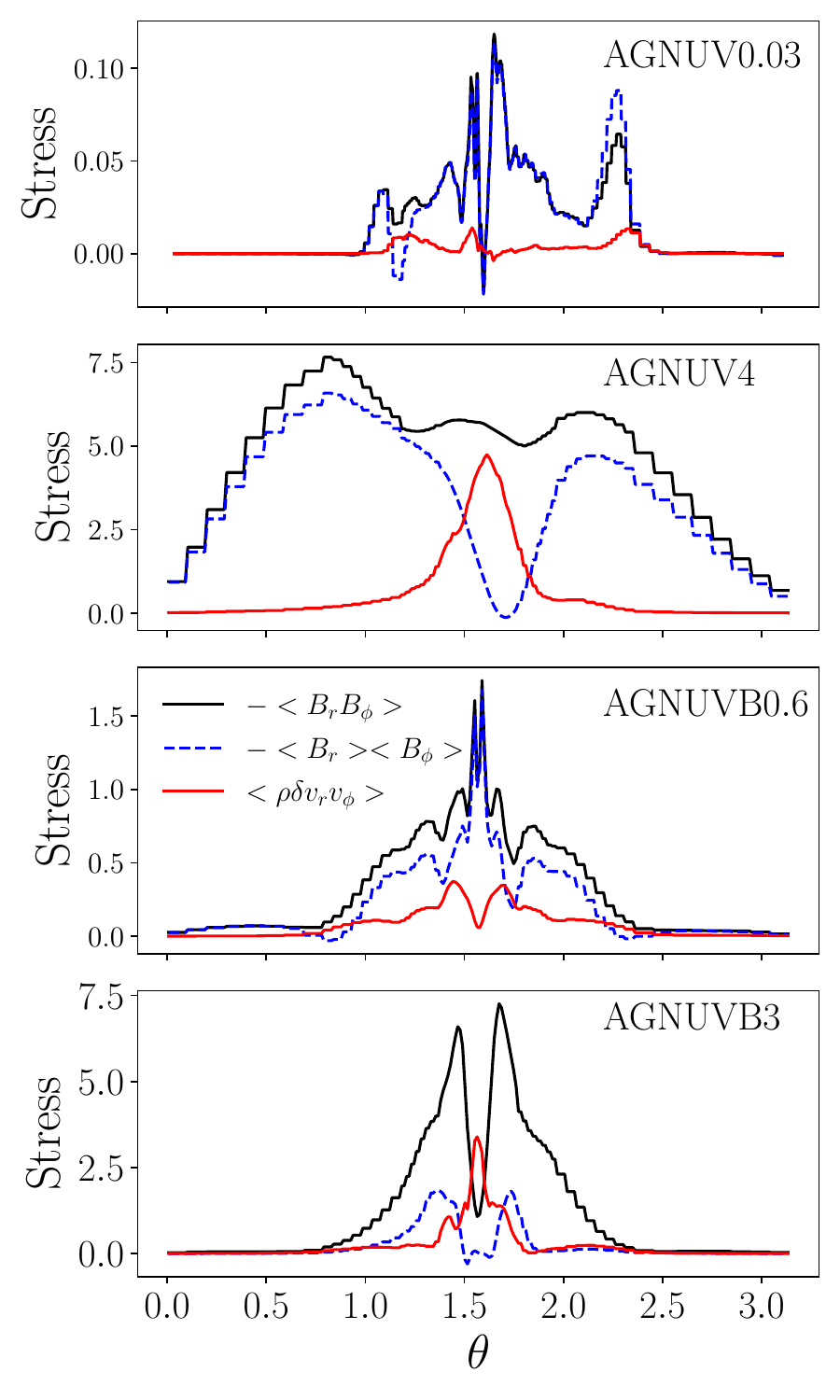}
	\caption{Time and azimuthally averaged vertical profiles of Maxwell stress (the black lines) and Reynolds stress (the red lines) at $r=100\rg$ for the four runs as indicated in each panel. The dashed blue lines are the Maxwell stress due to the azimuthally-averaged radial $\langle B_r\rangle$ and azimuthal $\langle B_{\phi} \rangle$ magnetic fields. The stress is in unit of $P_0$.}
	\label{vertical_stress}
\end{figure}

The time and azimuthally averaged density profiles at two representative radii $60\rg$ and $100\rg$ for the four runs are shown in Figure \ref{vertical_density}. For a fixed mass accretion rate, the midplane density increases with increasing radius. It also generally increases with increasing mass accretion rate for a fixed radius. However, the midplane density in run \sixteen\ is almost three orders of magnitude smaller compared with \eighteenB, which has a similar mass accretion rate. This is due to the efficient angular momentum removal by the magneto-centrifugal wind, which only shows up in run \sixteen.  

The ratios between density scale height and radius, as quantified by $\int_0^{\pi} \overline{\langle\rho\rangle} \theta d\theta/\int_0^{\pi} \overline{\langle\rho\rangle} d\theta$, are $\approx 2\%$ for \twelve, $\approx 3\%-4\%$ for \eighteenB, $\approx 5\%-6\%$ for \sixteenB, and always above $10\%$ for \sixteen. For disks with a similar accretion rate, the density scale height is always larger in the magnetic pressure-supported disk (\sixteen) compared with the scale height in the thermal pressure-supported disk (\eighteenB). Since magnetic pressure does not depend on density directly, hydrostatic equilibrium cannot constrain the density scale height for magnetic pressure-supported disks, which is very different from the  gas pressure-supported case. But the density scale height still generally decreases with decreasing mass accretion rate. This is very similar to the case of radiation pressure-supported disks, where the vertical density distribution is set by dissipation in the disk instead of hydrostatic equilibrium \citep{BLA11}.   

However, it is still unclear what sets the vertical density profiles in magnetic pressure-supported disks, despite the fact that such disks have been found in many simulations \citep{Gaburov+2012,Lanvov+2019,JIA19a,Jacquemin+2021}. It may depend on how MRI, Parker instability, and magnetic buoyancy operate together to set the final saturation state. Thermodynamics also plays an important role as the density scale height is closely related to the mass accretion rate, which determines the local cooling rate. Different assumptions have been made in the literature to relate the magnetic pressure in the saturation state to other disk quantities in the disk in order to construct accretion disk models in the magnetic pressure-dominated regime \citep{BegelmanPringle2007,BegelmanArmitage2023}. We find strong correlations between the Alfv\'en speed and sound speeds in the time and azimuthally averaged vertical structures of the disks in our four runs. We take $r=100r_g$ as a representative example and calculate the vertical profiles of Alfv\'en speed $\overline{\langle v_A\rangle}\equiv \sqrt{\overline{\langle P_B\rangle}/\overline{\langle \rho\rangle}}$, gas sound speed $\overline{\langle c_g\rangle}\equiv \sqrt{\overline{\langle P_g\rangle}/\overline{\langle \rho\rangle }}$, radiation sound speed $\overline{\langle c_r\rangle}\equiv \sqrt{\overline{\langle P_r\rangle }/\overline{\langle \rho\rangle }}$, as well as inflow speed $\overline{\langle v_r\rangle }\equiv \overline{\langle \rho v_r\rangle }/\overline{\langle \rho\rangle}$. As shown in Figure \ref{velocity_va_relation}, for the magnetic pressure-supported disks, $\overline{\langle c_r\rangle}$ and $\overline{\langle v_A\rangle}$ roughly follow a power law relation with the power law slope varying between $0.8$ and $1.3$. Similar relations also exist between $\overline{\langle c_g\rangle }$ and $\overline{\langle v_A\rangle}$ but with a smaller slope $\approx 0.4-0.7$.
For the run \eighteenB, there are clear changes of slopes between the radiation pressure supported midplane region and the upper layer where the Alfv\'en speed becomes comparable to or larger than the Keplerian speed $v_k$. Near the midplane, the Alfv\'en speed varies between $\approx 1\%$ and $10\%$ of the Keplerian speed. The inflow speed is typically smaller than $10^{-4}v_k$ near the midplane for the three runs \twelve, \sixteenB\ and \eighteenB. However, it can reach $0.1 v_k$ in run \sixteen\ with its strong magneto-centrifugal wind, which is also consistent with the smaller density for this run as shown in Figure \ref{vertical_density}. The scaling relationships between $v_A$, $c_g$ and $c_r$ we have identified here can be used to construct models of magnetic pressure supported accretion disks in future work.

\begin{figure}[htp]
	\centering
	\includegraphics[width=1.0\hsize]{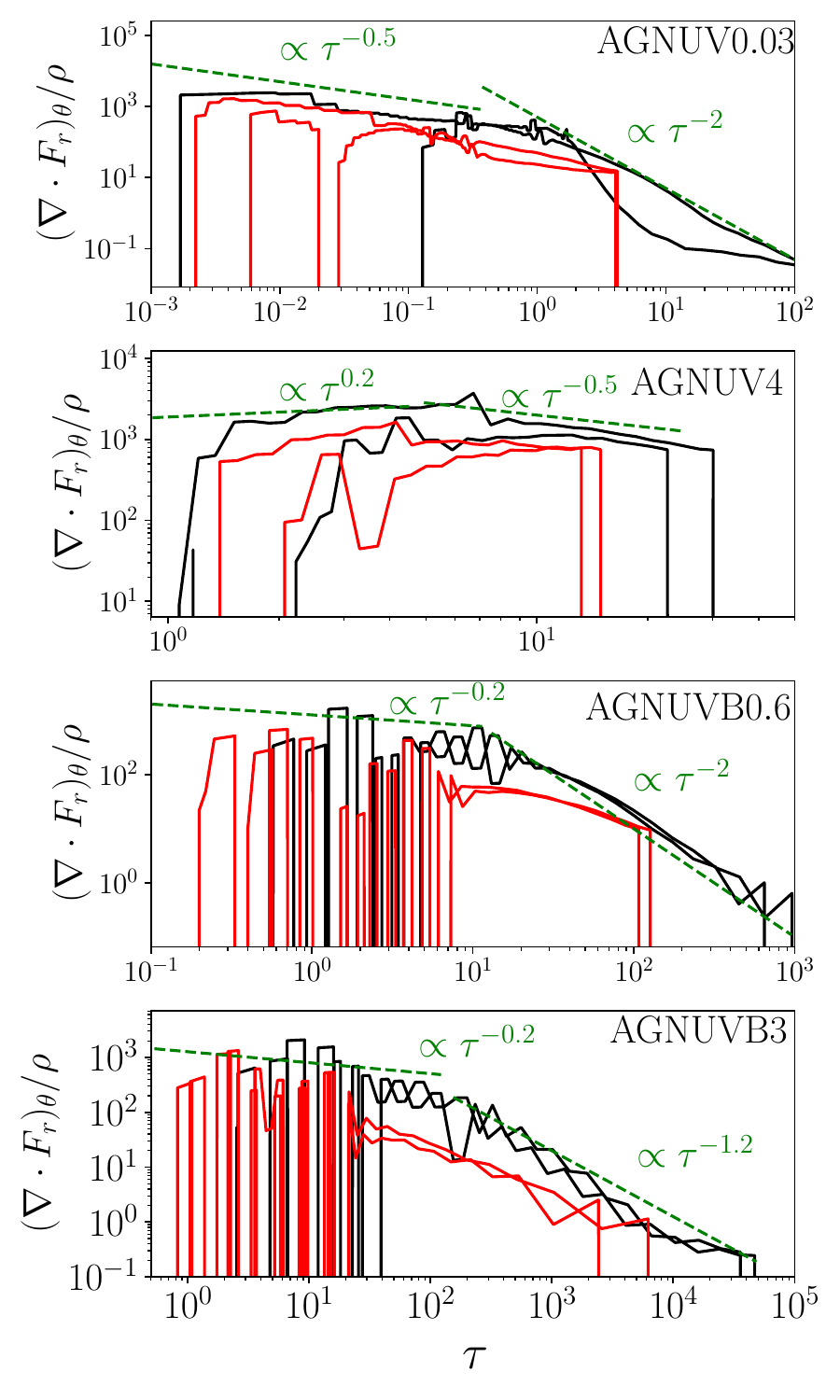}
	\caption{Dissipation, which is measured by the polar portion of the divergence of the azimuthally-averaged radiation flux $\left(1/r\sin\theta\right)\partial \left[\overline{\langle F_{\mathrm{r},\theta} \rangle}\sin\theta\right]/\partial \theta$,
    per unit mass, as a function of the integrated Rosseland mean optical depth from the rotation axis at radius $r=60r_g$ (red lines) and $r=100\rg$ (black lines) for the runs \twelve, \sixteen, \sixteenB, and \eighteenB, respectively (from top to bottom). The dashed green lines are power law scalings as indicated in each panel. We scale the flux divergence with the critical dissipation value $c\Omega_k^2/\kappa_{\rm es}$, where $\Omega_k$ is the Keplerian angular velocity at $r=100r_g$. Density is scaled with our fiducial unit $\rho_0=5\times 10^{-10}$ g/cm$^3$.}
	\label{vertical_dissipation}
\end{figure}

\subsection{Vertical Distributions of Dissipation}

The efficiency of angular momentum transport in the disks can be quantified by the Maxwell stress $-\overline{\langle B_rB_{\phi}\rangle}$ and Reynolds stress $\overline{\langle \rho \delta v_r v_{\phi} \rangle}$, where $\delta v_r \equiv v_r - \langle \rho v_r\rangle/\langle \rho\rangle$ is the radial velocity with the azimuthally averaged value subtracted for each snapshot in each cell. To distinguish Maxwell stress due to large-scale ordered magnetic fields, we also calculate the Maxwell stress due to the azimuthally averaged radial and azimuthal magnetic fields as $-\overline{\langle B_r\rangle \langle B_{\phi}\rangle}$. 

Vertical profiles of the stresses at the representative radius $r=100r_g$ for the four runs are shown in Figure \ref{vertical_stress}. The run with a radiation pressure supported disk \eighteenB\ shows that the Maxwell stress is dominated by the turbulent component and it peaks at a location away from the midplane. The Reynolds stress is strongly peaked at the midplane, due to the vertical density distribution. The total Maxwell stress integrated over the disk is $\approx 4$ times larger than the Reynolds stress. The ratio between the total stress and total thermal pressure is about $2-3\%$.  These are all consistent with the properties of MRI turbulence studied in the literature \citep{BLA11,Jiang+2013,Jiang+2016,Davis+2020}. For the two runs \twelve\ and \sixteenB\ where the disks are supported by magnetic pressure, $-\overline{\langle B_rB_{\phi}\rangle}$ is typically comparable to $-\overline{\langle B_r\rangle \langle B_{\phi}\rangle}$, which suggests that it is the large scale magnetic field that dominates the stress. For the case with a single loop of magnetic field initially in \twelve, the Maxwell stress reaches a minimum at the midplane. This is not the case for the run \sixteenB, which has two loops of magnetic fields initially. This magnetic topology can allow a net radial magnetic field at the midplane and thus non-zero Maxwell stress. The ratio of 
Maxwell stress to Reynolds stress is typically much larger than 
in \eighteenB. Even though $-\overline{\langle B_rB_{\phi}\rangle}$ is also different from $-\overline{\langle B_r\rangle \langle B_{\phi}\rangle}$ in run \sixteen, the difference is not due to small scale turbulence. Instead, it is mainly because the large-scale magnetic field is not strictly along the azimuthal direction and there are nonaxisymmetric perturbations in the location of the current sheet away from the equatorial plane. 
The ratio of total stress to thermal pressure varies from $\approx 0.3$ to $10$ for the three magnetic pressure-supported disks.

\begin{figure}[htp]
	\centering
	\includegraphics[width=1.0\hsize]{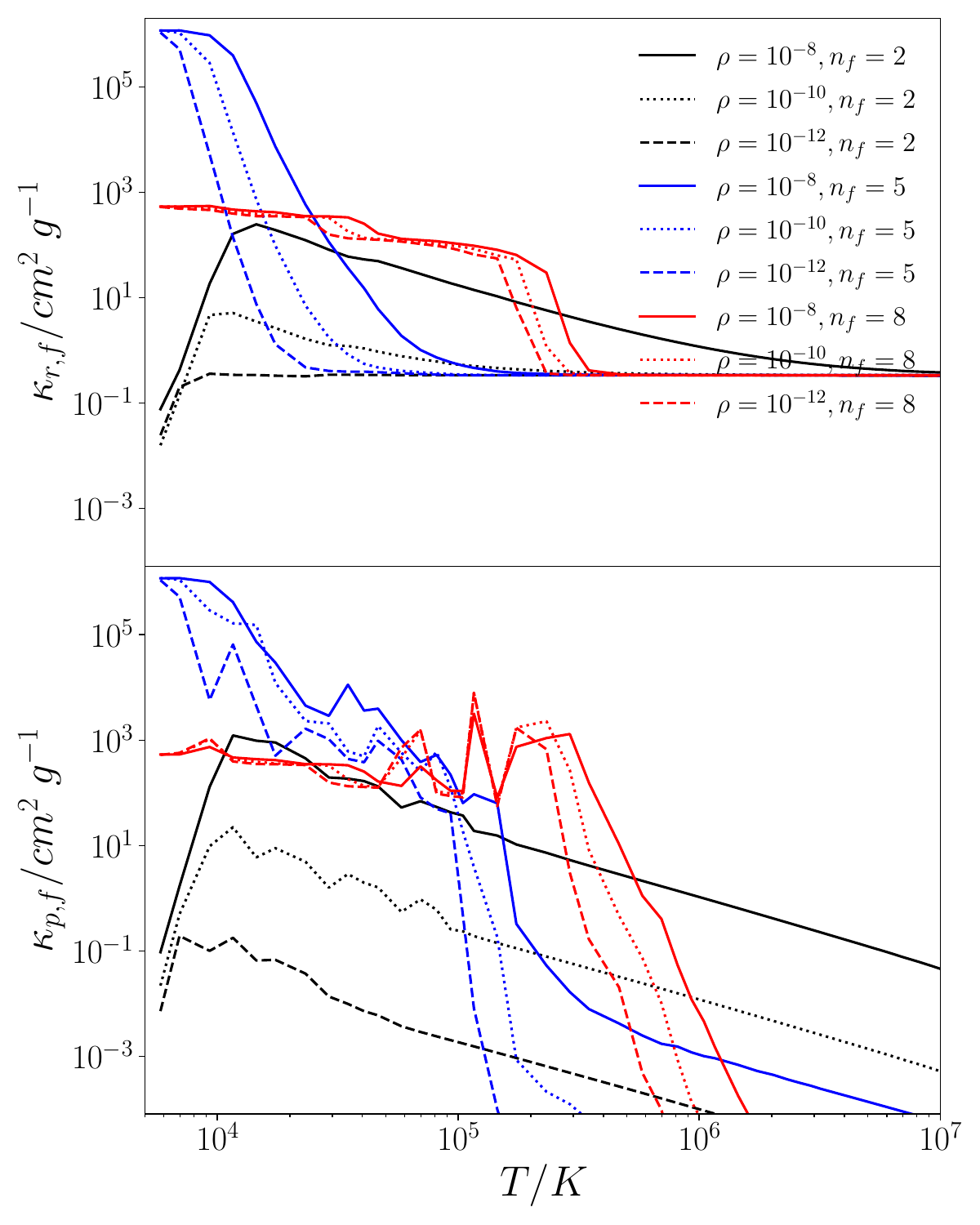}
	\caption{Rosseland mean (top panel) and Planck mean (bottom panel) opacities as a function of temperature for selected densities and frequency groups. The solid, dotted and dashed lines are for densities at $10^{-8}$ g/cm$^3$, $10^{-10}$ g/cm$^3$ and $10^{-12}$ g/cm$^3$ respectively. The black, blue and red lines are for frequency groups covering $h\nu/kT_0\in$ $[0.092,0.26]$, $[2.07,5.82]$, $[46.26,130.40]$ respectively.}
	\label{fre_opacity}
\end{figure}

Another important quantity that determines the vertical structures of accretion disks is the dissipation, which may not exactly follow the distribution of stress. In particular, the Maxwell stress contributed by the large scale mean magnetic fields does not necessary result in heating of the gas.  A good proxy for dissipation is the vertical spatial gradient of radiation flux, assuming heating of the gas is balanced by radiative cooling locally. We have checked that vertical velocities are small enough that vertical radiation flux is dominated by the diffusive flux. For radiation pressure supported disks, a critical dissipation value is $c\Omega_k^2/\kappa_{\rm es}$ if vertical gravity in the disk is balanced by the radiation force. Here $\Omega_k$ is the Keplerian angular velocity. Vertical flux gradients, scaled by this critical dissipation value, per unit mass, at $r=60r_g$ and $r=100r_g$ are shown in Figure \ref{vertical_dissipation} as a function of the Rosseland mean optical depth from the polar direction. Dissipation per unit mass typically increases with decreasing optical depth, as found by previous local shearing box simulations \citep{Krolik+2007,Jiangetal2014}. Near the midplane where most of the mass is, dissipation per unit mass roughly follows $\tau^{-\Gamma}$, where $\Gamma$ varies from $0.5$ in the run \sixteen\ to $2$ in \twelve\ and \sixteenB\ at $r=100r_g$. Near the surface, the slope is much flatter. For the same run, the slope is also slightly flatter at $r=60r_g$ compared with the values at $r=100r_g$.

\section{Spectrum Calculations}
\label{sec:spectrum}

\begin{figure*}[htp]
	\centering
	\includegraphics[width=0.45\hsize]{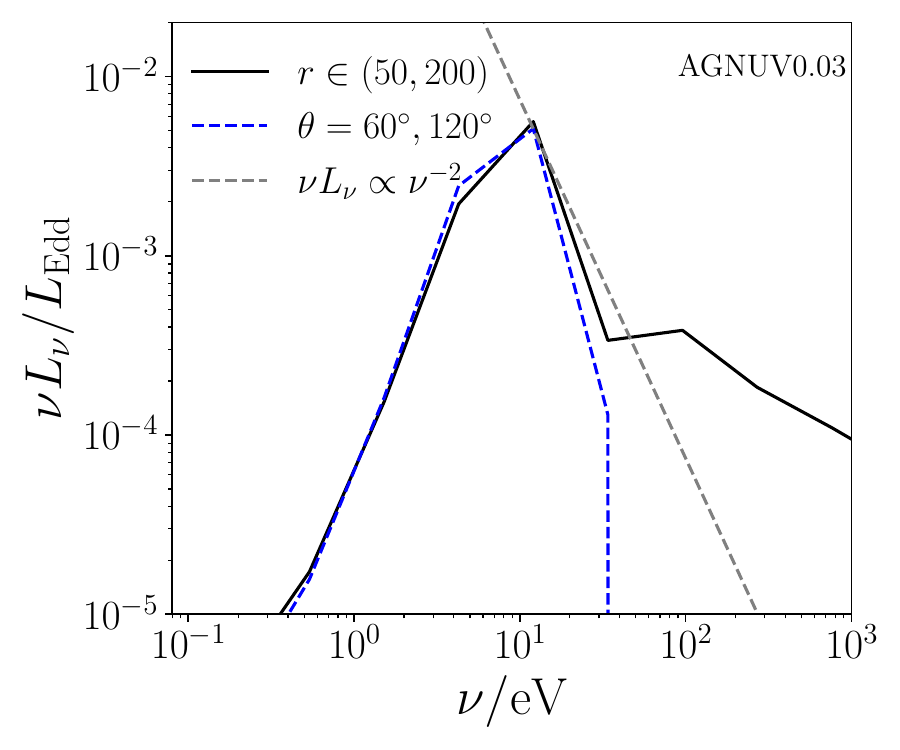}
	\includegraphics[width=0.45\hsize]{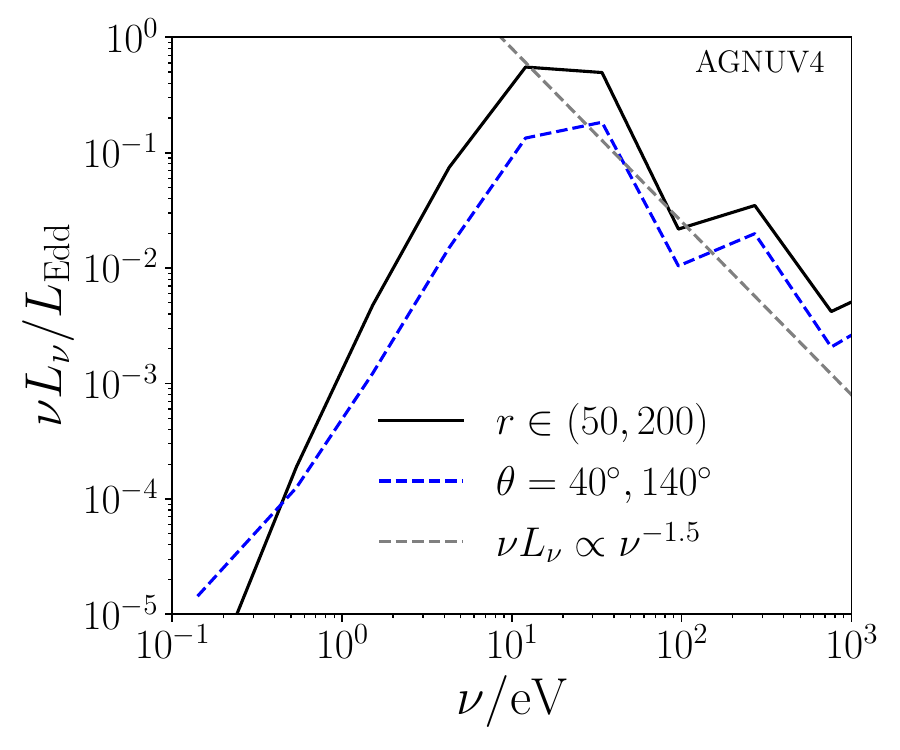}
        \includegraphics[width=0.45\hsize]{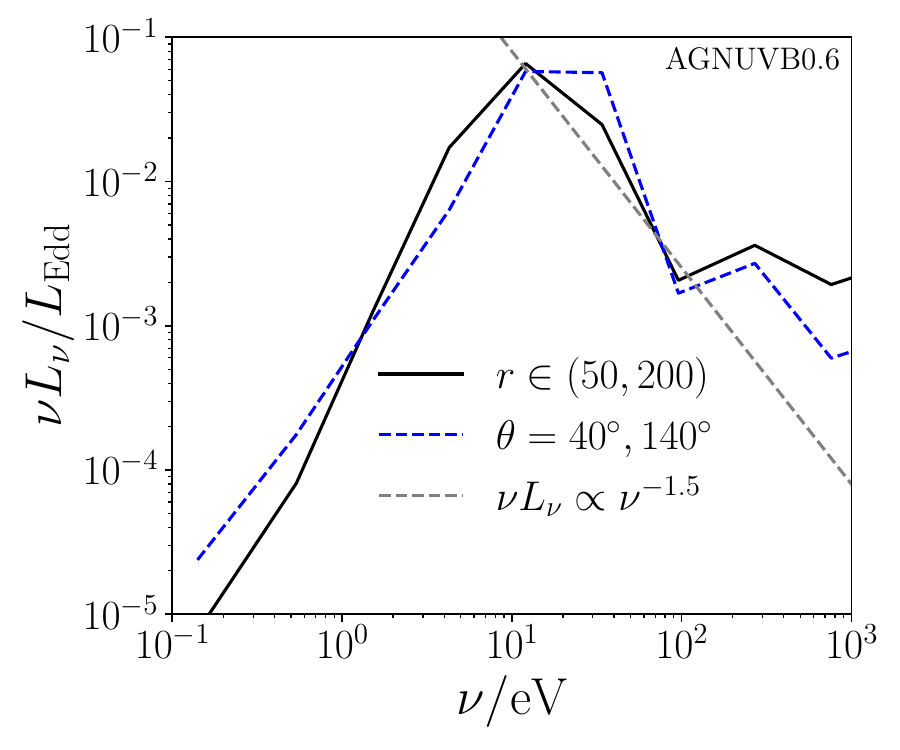}
	\includegraphics[width=0.45\hsize]{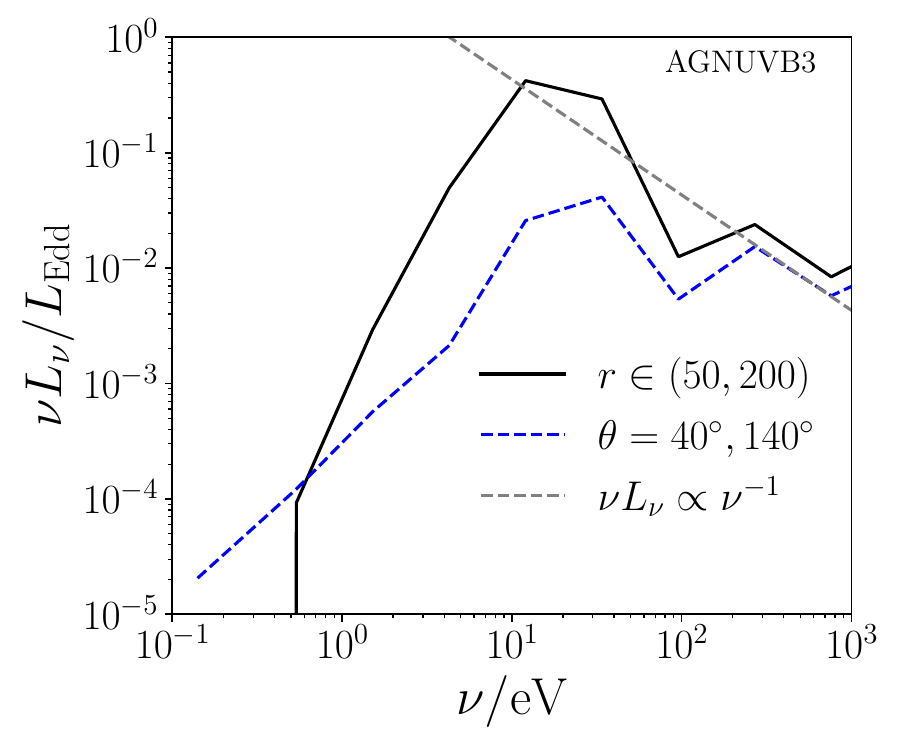}
	\caption{Radiative spectra for snapshots at times $4.26\times 10^6\ \rg/c$, $2.58\times 10^6\ \rg/c$, $3.87\times 10^6\ \rg/c$ and $2.71\times 10^6\ \rg/c$ of the four runs \twelve, \sixteen, \sixteenB\ and \eighteenB, respectively. The solid black lines are the results when we include all photons emitted between $50\rg$ and $200\rg$, while the dashed blue lines are the outcome if we only include photons leaving the surfaces at fixed $\theta$ values as indicated in each panel within $r<200\rg$. Emission from the polar region will therefore not be included in the dashed blue lines. The dashed gray lines are just power-law scalings to indicate the slopes of the spectra. 
  }
	\label{spec_collect}
\end{figure*}

We restart the frequency-integrated simulations described in previous 
sections from different snapshots using the multi-group radiation MHD 
module developed by \cite{Jiang2022} to calculate the broad-band spectra 
of the disks. We use 12 frequency groups covering 
$h\nu/kT_0\in[3.27\times 10^{-2},8.23\times 10^3)$ logarithmically. Two 
additional groups are used to cover $h\nu/kT_0\in[0,3.27\times 10^{-2})$ 
and $h\nu/kT_0\in[8.23\times 10^3,\infty)$, respectively. For each 
frequency group, we calculate the Rosseland mean and Planck mean 
opacities (as defined by \cite{Jiang2022} for each group) as a function of local gas temperature and density based on 
opacity tables provided by 
TOPS\footnote{\url{https://aphysics2.lanl.gov/apps/}} \citep{Colgan+2016} for solar metallicity. Examples of our adopted opacities as a function of temperature for three fixed densities in three representative frequency groups are shown in Figure \ref{fre_opacity}. As a consistency check, we note that the frequency-integrated Rosseland mean and Planck mean opacities provided by TOPS are also very similar to what we adopted in the original simulations. We emphasize that unlike many phenomenological models, there are no free parameters that we can adjust to control the spectra we get. They are just determined self-consistently by the dynamical structures of the disks we have simulated.

The cost to perform multi-group radiation transport calculations is linearly proportional to the number of groups we use. With 14 groups to resolve the broad-band spectra, it is too expensive to follow the dynamical evolution of the disks. Instead, we choose snapshots of the gray radiation MHD simulations when the disks have reached a steady state. Then we restart using 14 frequency groups by assuming the spectrum is a blackbody initially with the total radiation energy density matching the value in the gray calculations. We do not evolve the gas quantities and magnetic fields after the restart so that we can run the multi-group calculations to a steady state and get the spectra.

Spectra for snapshots of the four simulations are shown in Figure \ref{spec_collect}. We focus on the region inside $200\rg$ where steady-state disks have been established. We first calculate the total luminosity in each frequency group leaving the spherical surface at $r=200\rg$.  The resulting spectra are shown as the solid black lines in Figure \ref{spec_collect}. To ensure that photons emitted from the gas at larger radii do not contaminate the spectra we calculate, we also calculate the luminosity leaving a cylindrical region with radius $200\rg$ and varying heights from $150\rg$ to $250\rg$. The results are the same as long as the cylindrical height is above $200\rg$.
To test which part of the disk dominates the emission in the spectra, the total luminosity leaving spherical wedges at two fixed polar angles (one above the midplane and one below midplane) in each frequency group are shown as the dashed blue lines in Figure \ref{spec_collect}. The polar wedges are chosen to be $40^{\circ}$ and $140^{\circ}$ for \sixteen, \sixteenB\ and \eighteenB. For \twelve, we use $60^{\circ}$ and $120^{\circ}$ due to the smaller disk scale height. Compared with the solid black lines, spectra calculated this way will not include emission coming from the polar regions of the simulation domain.

\begin{figure*}[htp]
	\centering
         \includegraphics[width=0.45\hsize]{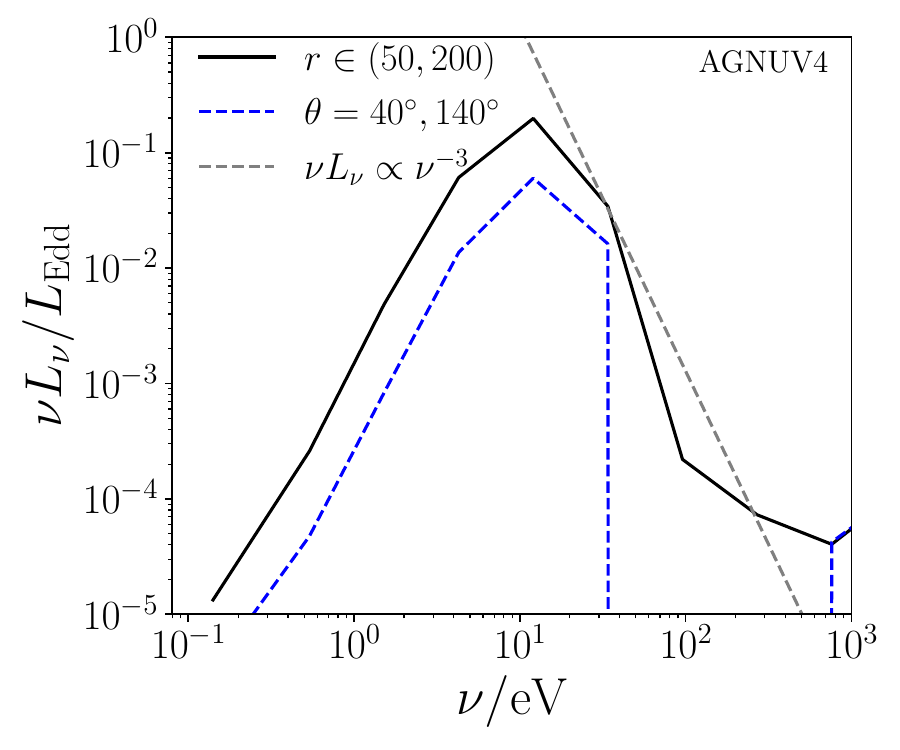}
	\includegraphics[width=0.45\hsize]{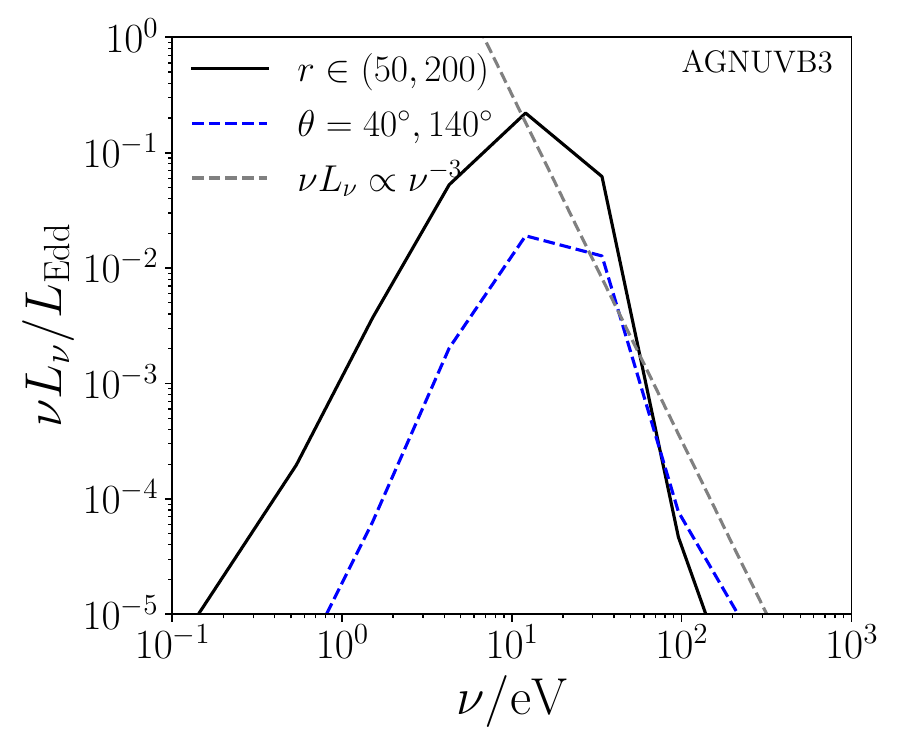}
	\caption{Spectra when the frequency shift due to the Doppler effect is turned off for the snapshot at time $2.58\times 10^6\ \rg/c$ of \text{AGNUV10} (left) and the snapshot at time $2.71\times 10^6\ \rg/c$ of \text{AGNUVB10} (right). All the lines represent the spectra calculated in the same way as shown in Figure \ref{spec_collect}.}
	\label{NoDoppler_spec}
\end{figure*}

All the spectra peak around $\sim 10$ eV, as the gas temperature at the photosphere for this region of the disk is $\approx 10^5$ K as shown in Figure \ref{wedge18B_16_Tgas}, and we have a limited range of temperature in the steady state region. The low-frequency part of the spectrum roughly follows a (Rayleigh-Jeans) blackbody shape. However, the high-frequency part of the spectrum has a power-law shape extending to $\sim 1$ keV, as shown in Figure \ref{spec_collect}. The spectral slopes vary from $\nu L_{\nu}\propto \nu^{-1}$ in \eighteenB\ to $\nu L_{\nu}\propto \nu^{-1.5}$ in 
\sixteen\ and \sixteenB. The power-law components show up in both the solid black lines and the dashed blue lines in these three simulation panels, showing
that they are produced by the main body of the disk instead of the polar region. This is not the case for run \twelve, which only has a weak power-law component if we include photons from the whole simulation box. The main body of the disk has a spectrum that falls off rapidly following a (Wien) blackbody shape for $\nu>20$ eV. 

The high-frequency power-law components cannot be produced by the thermal emission of the gas as the typical gas temperature in the disk is too small to produce these high-energy photons (cf. run \twelve). As an experiment, we turned off the thermal Compton term but it did not produce any difference as the typical gas temperature is too small compared with $m_ec^2/k_B$, where $m_e$ is the electron mass and $k_B$ is the Boltzmann constant. We then ran the multi-group calculations again but turned off the frequency shift due to the Doppler effect, which effectively turns off bulk Comptonization. The resulting spectra for the two runs \sixteen\  and \eighteenB\  are shown in Figure \ref{NoDoppler_spec}. The high energy power-law components have disappeared while the low frequency spectra remain the same. This strongly suggests that it is the bulk Comptonization that produces the power-law component. We have also tried to set all the velocities to zero and we get the same spectra as shown in Figure \ref{NoDoppler_spec}. 

Two types of bulk Comptonization have been discussed in the literature, which may be relevant for the spectrum formation in AGN disks. They all require the flow velocity to exceed the thermal velocity of electrons, which is $\approx  1.56\times 10^8$ cm/s for $T=10^5$ K, so that bulk Comptonization is more important compared with thermal Comptonization \citep{BlandfordPayne1981,PsaltisLambl1997}. This condition is easily satisfied for AGNs but not for X-ray binaries. One type of model attributes the super-thermal velocity to the small-scale turbulent velocity in the disk \citep{Thompson1994,Socratesetal2004,KaufmanBlaes2016,Kaufmanetal2018}, which can happen in AGN disks as radiation pressure is much larger than the gas pressure by a factor of $\approx 10^3-10^6$. Even if the turbulent kinetic energy density is limited to a small fraction of the radiation energy density, 
the turbulent velocity can still exceed the thermal speed of the electrons. 
At the scale of the \emph{ photon mean free path}, where photons are scattered by the turbulent eddies, their energies can be boosted to larger values compared to what thermal emission can normally produce. The slope of the photon spectrum produced by this mechanism depends on the effective Compton $y\equiv \left[c\rho\kappa_{\rm es}k_B T_{\rm eff}/\left(m_e c^2\right)\right]t_{\rm diff}$ parameter \citep{KaufmanBlaes2016}, where $T_{\rm eff}\propto m_e v^2/k_B$ is an effective temperature for electrons as related to the turbulent velocity and $t_{\rm diff}$ is the photon diffusion time across the characteristic length scale of the turbulence $l_v$. This Compton $y$ parameter is proportional to $\tau^2$, where $\tau$ is the optical depth across $l_v$. This mechanism is most effective in producing non-thermal photons when $y$ is in order of unity.

Another type of model proposes that non-thermal photons can be produced when the accretion disk has a coherent convergent flow \citep{PayneBlandford1981,Titarchuketal1997,Turollaetal2002}. The main difference compared with the previous type of bulk Comptonization is that the relevant velocity is not the small-scale turbulent velocity. Instead, it is the flow velocity at the scale of the disk. This mechanism also requires that the optical depth at the scale of the velocity variation is comparable to $c/v$ so that photons can be carried with the flow to experience the velocity gradient while they diffuse through the gas. In the standard thin disk model, the radial inflow speed is typically too small to make this mechanism work. But for 1D spherically symmetric super-Eddington accretion flow, the conditions can be met and the resulting radiation spectrum is determined by the velocity profile \citep{PayneBlandford1981}.


\begin{figure*}[htp]
	\centering
	\includegraphics[width=0.45\hsize]{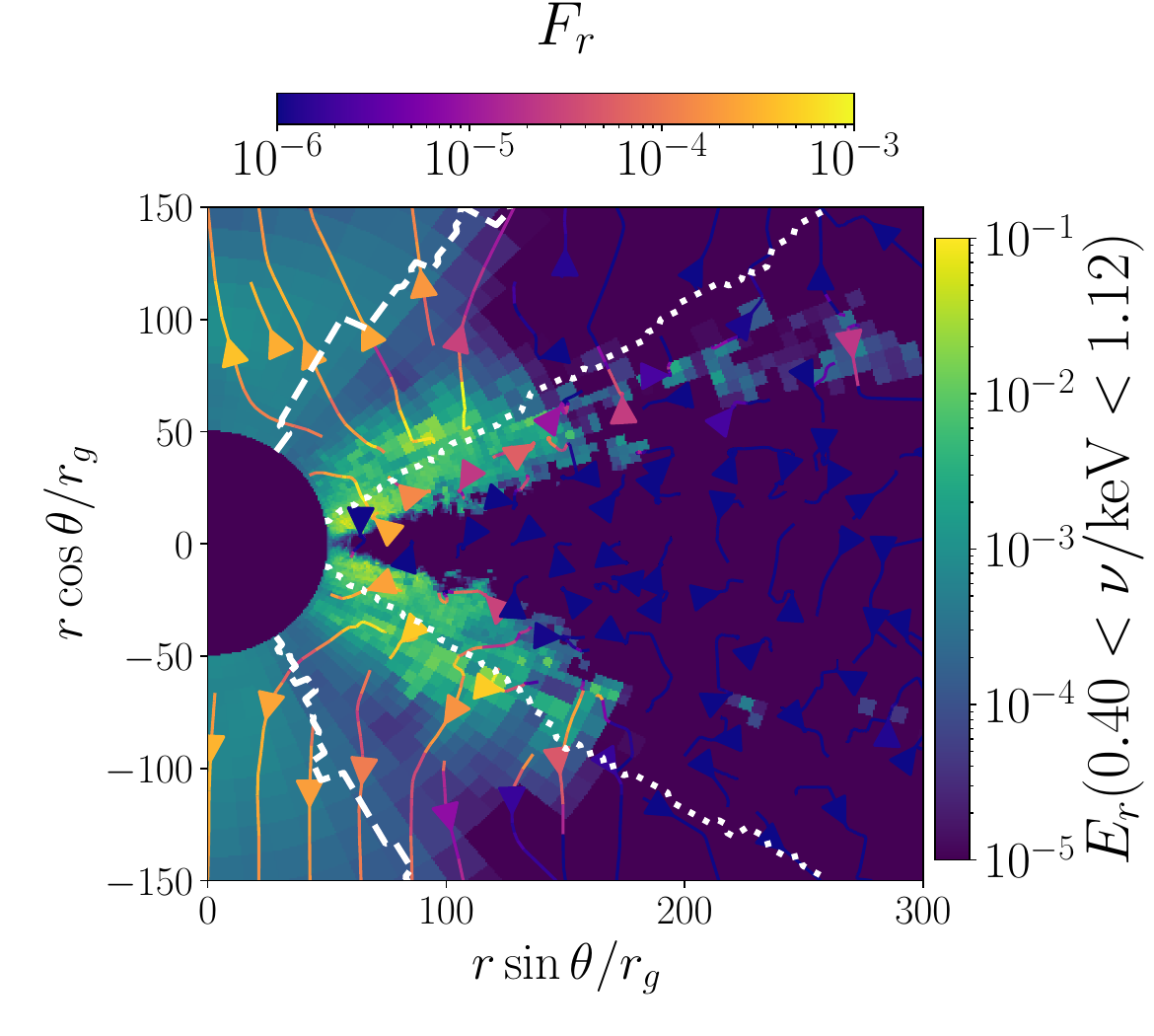}
	\includegraphics[width=0.45\hsize]{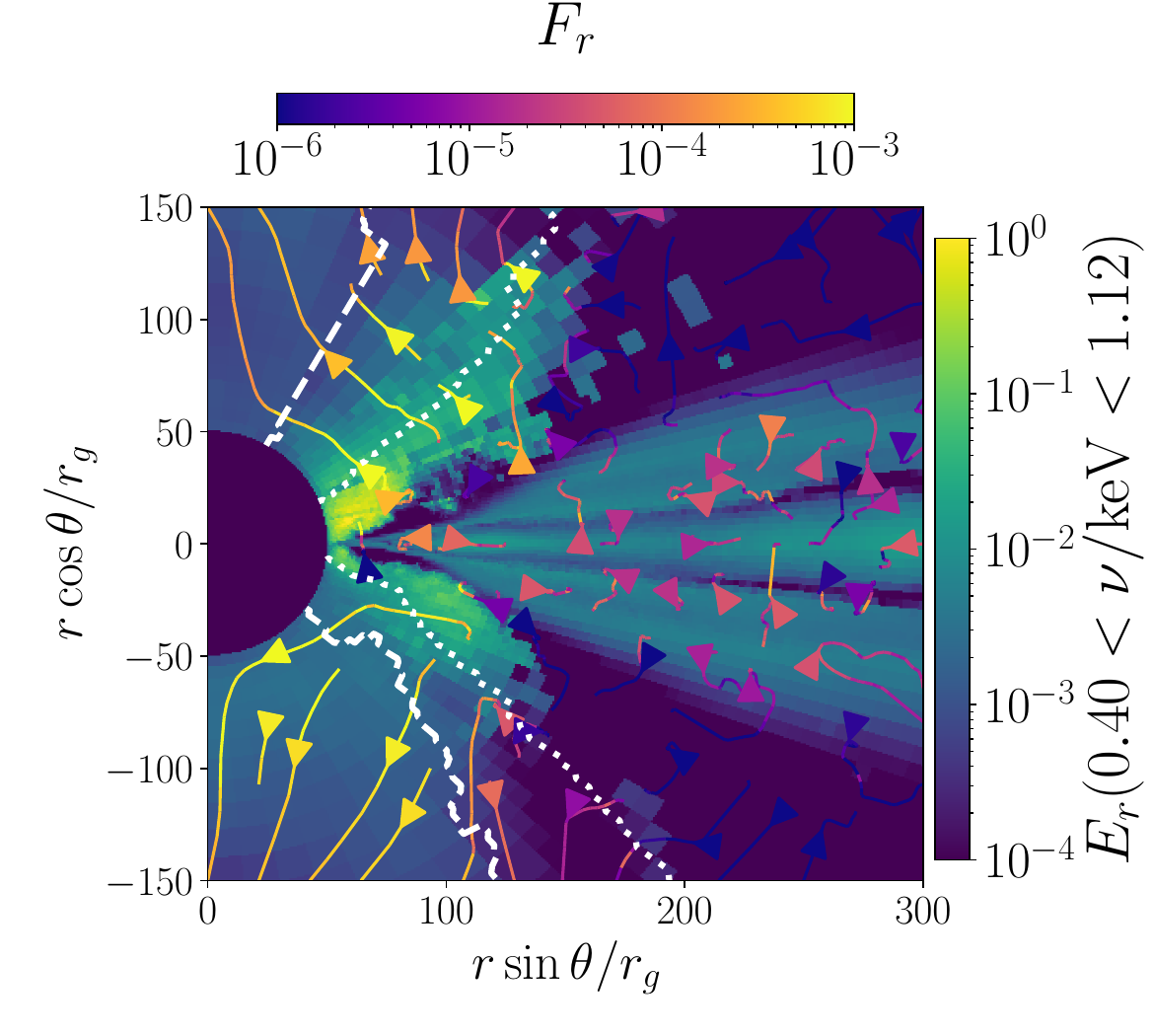}
	\caption{Spatial distribution of the azimuthally averaged 
    radiation energy density in the frequency group $0.40 < \nu/{\rm keV} < 1.12$ for the run \sixteenB\ (left) and \eighteenB\ (right). The streamlines are the radiation flux in the same frequency group. The dashed white lines indicate locations where the integrated Rosseland mean optical depth in this frequency group from the rotation axis is $1$ while the dotted white lines are the locations where the corresponding values are $10^3$.}
	\label{wedge18B_16B_Er_group}
\end{figure*}

\begin{figure}[htp]
	\centering
	\includegraphics[width=1.0\hsize]{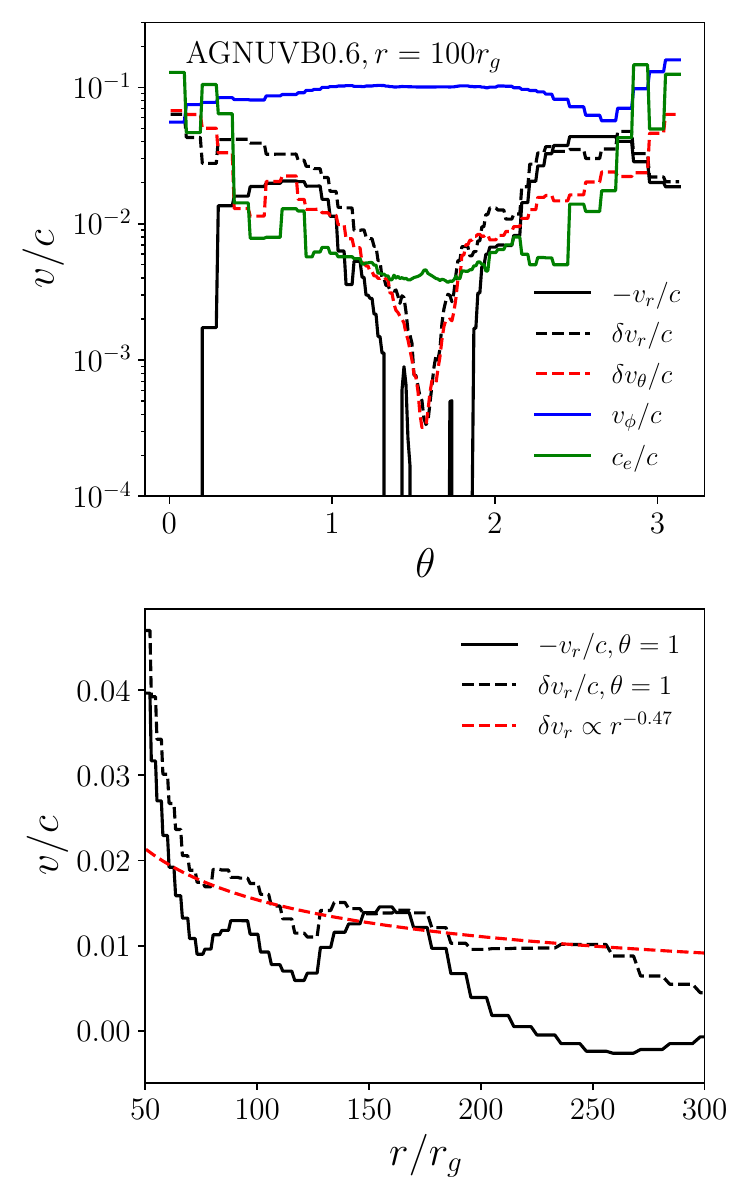}
	\caption{Top: Azimuthally averaged vertical profiles of different velocity components at radius $100\rg$ for the run \sixteenB\ at time $t=3.87\times 10^6\ \rg/c$. Bottom: Radial profiles of the azimuthally averaged radial velocity (solid black line for $-v_r$ and dashed black line for $\delta v_r$) at $\theta=1$ at the same time. 
    The dashed red line is just a power-law fit to $\delta v_r$. }
	\label{vertical_vel}
\end{figure}

To distinguish the physical mechanisms that are responsible for the soft X-ray photons in the disk, we can check where these photons are produced.
Spatial distributions of the azimuthally averaged radiation energy density and flux in the frequency group $(0.4<\nu/{\rm keV}<1.12)$ for the two selected snapshots of the run \sixteenB\ and \eighteenB\ are shown in Figure \ref{wedge18B_16B_Er_group}. Contributions to this frequency group from the main body of the disk are minimal, particularly for the run \sixteenB, because photons are mostly thermalized in this region of the disk and the gas temperature is too low to produce significant emission for this frequency range. Most of the high energy photons are produced away from the midplane with polar angle $30^{\circ} \lessapprox |\theta - 90^{\circ} |\lessapprox 50^{\circ}$,  where the gas has a much larger inflow speed compared with the gas in the disk midplane. This is also not in the optically thin region, as the integrated Rosseland mean optical depth from the rotation axis for this frequency group is $>10^3$, which is larger than $c/v_r$ as the typical inflow speed there is $v_r \gtrsim 0.1\% c$. Vertical profiles of the azimuthally averaged (weighted by density) radial ($v_r$) and rotational ($v_{\phi}$) velocities at $100\rg$ for the run \sixteenB\ at $t=3.87\times 10^6\ \rg/c$ are shown in Figure \ref{vertical_vel}. The rotational velocity is very close to the Keplerian value. 
We also show the vertical profiles of the turbulent radial ($\delta v_r$) and poloidal ($\delta v_{\theta}$) velocities, which are defined as 
\begin{eqnarray}
 \delta v_r \equiv \sqrt{\langle \rho v_r^2\rangle /\langle \rho\rangle},  \ \ 
 \delta v_{\theta} \equiv \sqrt{\langle \rho v_{\theta}^2\rangle /\langle \rho\rangle}.
\end{eqnarray}
They are comparable and both increase from $\approx 3\times 10^{-4}c$ at the midplane to $0.08c$ at the pole. As a comparison, the electron thermal speed, which is defined as $c_e\equiv \sqrt{1836\langle P_g\rangle/\langle \rho \rangle}$, is also shown as the green line in the top panel of Figure \ref{vertical_vel}. 
The radial velocity is smaller than $c_e$ at the midplane and becomes larger than $c_e$ when $|\theta-90^{\circ}| \gtrsim 30^{\circ}$. This is also the region where the high energy photons show up in Figure \ref{wedge18B_16B_Er_group}.

To quantify the radial profiles of radial velocities, we pick a fixed polar angle $\theta=1$ radian, where the radial velocity already exceeds the electron thermal velocity, and plot $-v_r$ and $\delta v_r$ as a function of radius in the bottom panel of Figure \ref{vertical_vel}. Excluding a rapidly rising region near the inner boundary, $\delta v_r$ roughly follows $r^{-0.47}$, which is not far from the radial dependence of the free-fall velocity, even though $\delta v_r$ is much smaller than the free-fall value. The spectral shape we get (Figure \ref{spec_collect}) is pretty close to the solution given by \cite{PayneBlandford1981} for their assumed free-fall velocity, which is also very close to the typical spectral shape observed for the soft X-ray excess.

\section{Discussion and Conclusions}
\label{sec:discussion}

Weak field MRI turbulence is evident in the radiation pressure dominated simulation \eighteenB, and it exhibits the hallmark toroidal field reversals in time that are the characteristic of the MRI butterfly diagram dynamo.  Such field reversals are not present in the magnetically dominated simulations, but some sort of dynamo activity is likely still present in runs  
\twelve\ and \sixteenB, as significant small scale turbulence exists. Indeed, some sort of dynamo enhancement of the magnetic fields over and above mere inward radial advection is necessary to explain the magnetic field pressure support that arises in the early evolution of the inner regions of run \sixteenB.  These two runs 
are very different from run \sixteen, which is completely dominated by the large scale, ordered magnetic fields. 

The physical mechanism that generates the turbulence in this regime is still unclear. It has been suggested that some convection or Parker-like instability caused by the buoyancy of the strong toroidal magnetic fields likely plays an important role \citep{ToutPringle1992,Gaburov+2012,Begelman+2022,Squire+2024}. The buoyantly rising magnetic field will have a vertical component locally, which is then sheared by the differential rotation of the disk. Non-axisymmetric MRI 
can also operate in this strongly magnetic pressure-dominated disk \citep{Pessah+2005,Das+2018}. It is often assumed that the Alfv\'en speed in the saturated state will reach $\sqrt{c_gv_k}$. As shown in Figure \ref{velocity_va_relation}, at $100\rg$, $v_A$ reaches $0.023v_k$ for the run \twelve\ and $\approx 0.056 v_k$ for the run \sixteenB\ near the midplane, while $c_g$ and $c_r$ are $ 3.74\times 10^{-4}v_k, 1.05\times 10^{-3}v_k$ for the run \twelve\ and $8.81\times 10^{-4}v_K, 0.01 v_k$ for the run \sixteenB\ at the same location. These results are not very far from the assumed saturation state with $v_A$ different by $20\%-50\%$. However, this assumption does not work for the run \sixteen. This is perhaps not surprising as turbulence did not develop in the magnetically dominated regions for this run. 

The role of cooling in the dynamo cycle is unexplored in the literature, which likely also plays an important role. The balance between heating and cooling will determine the density scale height, even in the magnetic pressure dominated disks. Smaller density scale height typically means smaller inflow speed, which will increase the timescale for radial transport of magnetic flux. Detailed studies on the dynamo process for disks with different accretion rates will be carried in the future.

The simulations are designed to cover the radial range between $50\rg$ and $200\rg$, where the UV photons and soft X-ray exccess are expected to be produced. We do not include the region inside $50\rg$ to avoid very small time steps. Simulations for AGN accretion disks covering the inner region \citep{JIA19a,JIA19b} show that gas temperature will increase with decreasing radius as expected and X-ray coronae can be formed inside $\approx 20\rg$. The funnel region in the four simulations shown here will be significantly impacted by the radiation emitted from the inner region. However, the optically thick part of the disks, where soft X-ray photons are produced, will likely not be affected as hard X-rays will not be able to penetrate this part of the disk \citep{Secunda+2024,Secunda+2025}. 

Since bulk Comptonization boosts thermal photons to higher energies in a continuous way, the power law component that we find smoothly connects to the thermal peak of the disk at $\sim10$~eV.  This is tantalizingly close to the observed 1000\AA\ (12 eV) break observed in quasars \citep{ZHE97,LAO97}.  This might be due to opacity effects (e.g. the hydrogen Lyman edge) in our multi-frequency group calculation, but it also might simply be due to the fact that the characteristic temperature of our disks is $10^5$~K ($\sim10$~eV) for our chosen $10^8\msun$ black hole mass.  More work needs to be done with simulations of different black hole masses and additional frequency groups to understand the connection between the thermal and power-law components of the spectrum.  For smaller mass black holes, as in Narrow Line Seyfert 1 galaxies, we expect that hydrogen will be more ionized and opacities will be reduced, and the whole spectrum will likely move to higher frequency.  However, the shape of the power law will likely remain the same as this is due to the physics of optically thick bulk Comptonization.


All four simulations described here are optically thick near the midplane, even those that are supported by magnetic pressure. This is necessary to keep the gas and radiation thermalized so that the photons emitted by the disk are in the same frequency range as what is observed. This is also one of the fundamental assumptions in the standard thin disk model, even though this model fails to explain many observational properties of AGNs. But the radiation spectrum does put strong constraints on new accretion disk models. For example, if accretion disks have inflow speeds comparable to the free-fall speed as suggested by \cite{Hopkins+2024}, the effective vertical optical depth across the disk will be smaller than unity for sub-Eddington disks.  Indeed, this is evident in Figure 7 of \cite{Hopkins2024}. Gas and radiation will then \emph{not} be thermalized in this case and the radiation spectrum produced by such a disk will likely be very different from what is observed.

In summary, the main results we find based on our numerical experiments are:
\begin{itemize}
    \item Accretion disks with lower accretion rates
    tend to have a structure in which vertical gravity in the disk is balanced by strong toroidal magnetic fields.  Disks with higher accretion rates can be dominated by radiation pressure.  In simulations, this occurs provided the cooling time scale of the gas is longer than the timescale it takes for the turbulence to develop and produce the early inflow of gas.  After that, the flow is self-sustaining.
    \item The vertical density distribution in the disk is determined by cooling and dissipation in the disk, for both magnetic pressure and radiation pressure supported disks. Most of the mass is concentrated near the midplane with the mass weighted inflow speed typically smaller than $10^{-4}-10^{-3}$ of the Keplerian speed except for the case where strong magneto-centrifugal winds are developed. In that case, the inflow speed can reach $10\%$ of the Keplerian speed. The density scale height decreases with decreasing accretion rate, becoming $1\%-3\%$ of the radius for the disk with $3\%$ Eddington accretion rate. This is likely unresolved in galaxy-scale simulations, but the dense optically thick gas is important for producing the thermal emission that is observed in AGNs. Therefore the radiation spectrum is a key way of testing these simulation results.
    \item A significant fraction of accretion happens in a region away from the midplane for the magnetic pressure supported disks. However, most of this region is still very optically thick for disks with a relatively high accretion rate. Only when the accretion rate is $\approx 3\%$ of the Eddington accretion rate does most of the accretion happen in in an optically thin region. 
    \item Magnetic pressure supported disks with net poloidal magnetic fields through the disk do not always generate magneto-centrifugal winds, as is commonly found in isothermal simulations with net poloidal magnetic fields \citep{BAI13,Gressel+2015,Jacquemin+2021}.  Comparing the two runs \twelve\ and \sixteen, the wind only develops when the accretion rate is high and the cooling timescale is not too short so that relatively large disk scale heights can be maintained. 

    \item In magnetic pressure supported AGN accretion disks, the Alfv\'en speed has strong correlations between the radiation sound speed $c_r$ and gas sound speed $c_g$ vertically as $c_r\propto v_A^{0.8-1.3}$, $c_g\propto v_A^{0.4-0.7}$. The radiation sound speed is typically larger than the gas sound speed by one to two orders of magnitude. Combined with the vertical dissipation profiles we get from the simulations, this can be used to construct physical models of magnetic pressure supported disks.

    \item For disks with accretion rates close to the Eddington value, power law spectra varying between $\nu^{-1}$ and $\nu^{-2}$ can be formed in the frequency range $\approx 0.01-1$ keV for the $10^8\msun$ black hole we study here. These slopes are comparable to what is observed \citep{LAO97}. 
    This non-thermal spectrum is formed due to bulk Comptonization of the optically thick inflowing gas in the disk just above the midplane.
\end{itemize}

\acknowledgments
We thank Chicuan Jin and Chris Done for helpful discussions during the Aspen Workshop on Accretion Physics in the Era of JWST, where progress was made on this work. We also thank Julian Krolik and the referee for helpful comments that improved the paper. The Aspen Center for Physics is supported by National Science Foundation grant PHY-2210452.  This work was also supported in part by NASA Astrophysics Theory Program grant 80NSSC22K0820, and
the TCAN collaboration funded by the grant 80NSSC21K0496.
Resources supporting this work were provided by the NASA High-End Computing (HEC) Program through the NASA Advanced Supercomputing (NAS) Division at Ames Research Center. The Center for Computational Astrophysics at the Flatiron Institute is supported by the Simons Foundation.

\bibliographystyle{aasjournal}
\bibliography{citations}

\end{CJK*}

\end{document}